\documentclass[twocolumn,trackchanges]{aastex701}
\usepackage{graphicx}
\usepackage{float}
\usepackage{amsmath,amssymb}
\usepackage{xcolor}
\usepackage{multirow} 

\def\swift{{\it Swift}}

\def\TDE{TDE\,2022dbl}

\usepackage{colortbl}

\shorttitle{\TDE}
\shortauthors{Guo et al.}
\graphicspath{{./}{figures/}}
\begin{document}

\title{Seeing Double: The Theory Defying Twin Flares of \TDE}

\author[0009-0002-9727-8326]{Xinze~Guo}
\affiliation{Center for Astrophysics $|$ Harvard \& Smithsonian, 60 Garden Street, Cambridge, MA 02138--1516, USA}
\affiliation{Department of Astronomy, University of California, Berkeley, CA 94720-3411, USA}
\email{xinze.guo@cfa.harvard.edu}

\author[0000-0002-7706-5668]{Ryan~Chornock}
\affiliation{Department of Astronomy, University of California, Berkeley, CA 94720-3411, USA}
\affiliation{Berkeley Center for Multi-messenger Research on Astrophysical Transients and Outreach (Multi-RAPTOR), University of California, Berkeley, CA 94720-3411, USA}
\email{chornock@berkeley.edu}

\author[0000-0003-4768-7586]{Raffaella Margutti}
\affiliation{Department of Astronomy, University of California, Berkeley, CA 94720-3411, USA}
\affiliation{Berkeley Center for Multi-messenger Research on Astrophysical Transients and Outreach (Multi-RAPTOR), University of California, Berkeley, CA 94720-3411, USA}
\affiliation{Department of Physics, University of California, 366 Physics North MC 7300, Berkeley, CA 94720, USA}
\email{rmargutti@berkeley.edu}

\author[0000-0001-6747-8509]{Yuhan Yao}
\affiliation{Department of Astronomy, University of California, Berkeley, CA 94720-3411, USA}
\affiliation{Berkeley Center for Multi-messenger Research on Astrophysical Transients and Outreach (Multi-RAPTOR), University of California, Berkeley, CA 94720-3411, USA}
\affiliation{Miller Institute for Basic Research in Science, 206B Stanley Hall, Berkeley, CA94720, USA}
\email{yyao@pku.edu.cn}

\author[0000-0002-5698-8703]{Erica Hammerstein}
\affiliation{Department of Astronomy, University of California, Berkeley, CA 94720-3411, USA}
\affiliation{Berkeley Center for Multi-messenger Research on Astrophysical Transients and Outreach (Multi-RAPTOR), University of California, Berkeley, CA 94720-3411, USA}
\email{ekhammer@berkeley.edu}

\author[0000-0002-2249-0595]{Natalie LeBaron}
\affiliation{Department of Astronomy, University of California, Berkeley, CA 94720-3411, USA}
\affiliation{Berkeley Center for Multi-messenger Research on Astrophysical Transients and Outreach (Multi-RAPTOR), University of California, Berkeley, CA 94720-3411, USA}
\email{nlebaron@berkeley.edu}

\author[0000-0003-1673-970X]{S.~Bradley Cenko}
\affiliation{Astrophysics Science Division, NASA Goddard Space Flight Center, Greenbelt, MD 20771, USA}
\affiliation{Joint Space-Science Institute, University of Maryland, College Park, MD 20742, USA}
\email{brad.cenko@nasa.gov}

\author[0000-0002-3168-0139]{Matthew J. Graham}
\affiliation{Cahill Center for Astronomy and Astrophysics, California Institute of Technology, Pasadena, CA 91125, USA}
\email{mjg@caltech.edu}

\author[0009-0008-0782-5028]{Dina Ibrahimzade}
\affiliation{Department of Astronomy, University of California, Berkeley, CA 94720-3411, USA}
\email{dinaevazade@berkeley.edu}

\author[0000-0003-0466-3779]{Itai Sfaradi}
\affiliation{Department of Astronomy, University of California, Berkeley, CA 94720-3411, USA}
\affiliation{Berkeley Center for Multi-messenger Research on Astrophysical Transients and Outreach (Multi-RAPTOR), University of California, Berkeley, CA 94720-3411, USA}
\email{itai.sfaradi@berkeley.edu}

\author[0000-0001-8426-5732]{Jean Somalwar}
\affiliation{Department of Astronomy, University of California, Berkeley, CA 94720-3411, USA}
\affiliation{Berkeley Center for Multi-messenger Research on Astrophysical Transients and Outreach (Multi-RAPTOR), University of California, Berkeley, CA 94720-3411, USA}
\affiliation{Kavli Institute for Particle Astrophysics and Cosmology, Stanford, CA 94305, USA}
\email{jsomalwar@berkeley.edu}

\author[0000-0003-2434-0387]{Robert Stein}
\affiliation{Department of Astronomy, University of Maryland, College Park, MD 20742, USA}
\affiliation{Joint Space-Science Institute, University of Maryland, College Park, MD 20742, USA}
\affiliation{Astrophysics Science Division, NASA Goddard Space Flight Center, Mail Code 661, Greenbelt, MD 20771, USA}
\email{rdstein@umd.edu}

\author[0009-0002-4843-2913]{Eli Wiston}
\affiliation{Department of Astronomy, University of California, Berkeley, CA 94720-3411, USA}
\affiliation{Berkeley Center for Multi-messenger Research on Astrophysical Transients and Outreach (Multi-RAPTOR), University of California, Berkeley, CA 94720-3411, USA}
\email{ewiston@berkeley.edu}

\author[0000-0002-0987-3372]{Joahan Castaneda Jaimes}
\affiliation{Division of Physics, Mathematics and Astronomy, California Institute of Technology, 1200 E. California Blvd, Pasadena, CA 91125, USA}
\email{jocastan@ipac.caltech.edu}
    
\author[0000-0002-5619-4938]{Mansi M. Kasliwal}
\affil{Division of Physics, Mathematics, and Astronomy, California Institute of Technology, Pasadena, CA 91125, USA}
\email{mansi@astro.caltech.edu}

\author[0000-0003-2451-5482]{Russ R. Laher}
\affiliation{IPAC, California Institute of Technology, 1200 E. California Blvd, Pasadena, CA 91125, USA}
\email{laher@ipac.caltech.edu}

\begin{abstract}
Partial tidal disruption events, in which a star survives a close passage by a supermassive black hole and may be disrupted again on a subsequent encounter, provide a unique probe of supermassive black hole properties and accretion physics. We examine TDE\,2022dbl, a partial tidal disruption event candidate that exhibits two separate UV/optical flares, using 11 new optical spectra, updated UV/optical photometry, and a Keck-II/ESI host-galaxy spectrum. We trace the evolution of the prominent blended \ion{N}{3}+\ion{He}{2} feature across both flares and find that the blend evolves from \ion{N}{3}-dominated to \ion{He}{2}-dominated emission, which may reflect changing ionization conditions and a declining efficiency of Bowen fluorescence. The two flares show similar thermal evolution and spectral features, supporting a partial-disruption scenario. The second flare exhibits a broader, plateau-like peak, a longer decline timescale, and less extreme line velocities than the first flare. We suggest that these differences arise from the interaction between debris from the second disruption and an accretion disk formed during the first flare. From the host-galaxy velocity dispersion, we infer a black hole mass of $\log_{10}(M_{\rm BH}/M_\odot)=6.01\pm0.45$. No third flare has been detected at the expected recurrence epoch, leading us to propose a new interpretation in which a positive-energy kick during ballistic core reformation delays the return of the surviving remnant.
\end{abstract}



\section{Introduction} \label{sec:intro} 
A tidal disruption event (TDE) occurs when a star travels close to the tidal radius ($r_t$) of a super-massive black hole (SMBH) in the center of its host galaxy and is ripped apart. Roughly 50\% of the stellar debris is accreted onto the SMBH, producing a flare of electromagnetic emission that temporarily lights up the otherwise dormant SMBH \citep{rees1988tidal, evans1989tidal}. TDEs are therefore critical probes for detecting and studying dormant SMBH properties, such as mass \citep{mockler2019weighing} and spin \citep{wen2020continuum}, and offer a unique opportunity to study accretion physics. TDEs are observed across a broad range of wavelengths, with their observed emission depending on viewing angle and reprocessing by optically thick debris \citep{Guillochon2014Provided, roth2016x, dai2018unified}. Although optically-selected TDEs have an infrequent occurrence rate of approximately $10^{-5}$ -- $10^{-4}$ per galaxy per year \citep{yao2023tidal}, large time-domain surveys have discovered over 100 TDEs from radio to $\gamma$-ray wavelengths \citep{gezari2021tidal, hammerstein2022final, yao2023tidal, Langis2026TDEs, franz2026open, Zhang2026}.

When a star only grazes the SMBH, corresponding to a penetration factor $\beta = \frac{r_t}{r_p} < 1$, where $r_p$ is the pericenter distance, it may not be completely disrupted during its first close passage and can return for subsequent encounters. These events are called partial TDEs (pTDEs), in contrast to full TDEs (fTDEs), where the star is completely disrupted during the first encounter. pTDEs are expected to reside in galactic nuclei with inefficient loss-cone refilling, where stars move inward more slowly, allowing them to reach the pericenter range for partial disruption before complete disruption \citep{bortolas2023partial, Broggi2024pTDE}. The rate of pTDEs is predicted to be higher than that of fTDEs \citep{stone2020rates, zhong2022revisit, bortolas2023partial}, yet observationally confirmed examples remain rare. This scarcity makes well-observed pTDE candidates especially valuable for constraining the underlying physics.

Several nuclear transients have been proposed as pTDE candidates, but their nature and observational coverage vary substantially. TDE\,2018fyk showed rebrightening in X-ray and UV $\sim1200$ days after the first peak, following a complex first flare with a double-peaked light curve and iron-rich spectra \citep{wevers2019evidence, wevers2023live}. TDE\,2020vdq rebrightened in the optical/UV $\sim947$ days after discovery, although its first flare lacks comparable UV and spectroscopic coverage \citep{somalwar2023first}. ASASSN-14ko has exhibited $>$20 periodic optical flares every $\sim$114 days \citep{payne2021asassn, payne2023chandra}, but its spectra are AGN-like, and alternative explanations, such as repeated star–disk interactions, make its origin uncertain \citep{linial2024period}. AT\,2023uqm has been observed with five flares every $\sim$526.75 days with an approximately exponential increase in energy \citep{wang2025star}, with the first four flares lacking consistent UV/X-ray coverage.

TDE\,2022dbl is one of the best-observed pTDE candidates to date, with two separate outbursts peaking in February 2022 and February 2024 \citep{arcavi20222018mac, yao2024ztf}. It has an unusually rich, well-sampled, multi-wavelength dataset, and both flares were spectroscopically classified as TDEs \citep{lin2024unluckiest, hinkle2024double, makrygianni2025double}. No significant X-ray emission is detected during the first and second flares, with one weak detection between the flares, at 371 days after the first peak \citep{hinkle2024double}. The two flares share comparable UV/optical luminosities at the lower end of the TDE population. Faint radio emission is detected at 167 and 519 days after the first peak, while no clear detection is associated with the second flare \citep{makrygianni2025double}. TDE\,2022dbl is identified as a Bowen TDE based on the presence of \ion{N}{3} emission lines. However, a detailed, systematic analysis of the evolution of individual spectral features, particularly Bowen emission features, remains limited. 

Moreover, TDE\,2022dbl was predicted to produce a third flare near February 2026 if its recurrence remained periodic, but no comparable third flare has been observed. The late-time emission around this epoch has a bolometric luminosity of only $1.1^{+1.2}_{-0.8}\times10^{42}$ erg s$^{-1}$, approximately 30 times fainter than the second flare \citep{hinkle2026disappearingactconstraintsmissing}. That study found that two independent TDEs are statistically unlikely and that a binary double-TDE scenario is inconsistent with the observed recurrence timescale, instead favoring a twin-flare pTDE involving a main-sequence star on a bound orbit after a deep initial encounter. Here, we explore an additional interpretation in which a positive-energy kick during ballistic core reformation places the surviving stellar remnant on a longer-period orbit, delaying its return to pericenter \citep{coughlin2025s}.

In this work, we investigate TDE\,2022dbl using an improved and expanded analysis aimed at resolving several questions about its repeated flaring behavior. We present a revised reduction of the UV/optical photometry, with the UV analysis explicitly accounting for instrumental systematics, including image trailing and PSF effects, that were not considered in previous studies. We derive an independent SMBH mass estimate from host galaxy modeling using our late-time high-resolution spectrum. Taking advantage of our extensive spectroscopic sequence, we trace the evolution of spectral features across both flares. By comparing both flares, we investigate the physical origin of their photometric and spectroscopic differences and explore an additional interpretation for the missing third flare. The structure of this work is as follows. In Section~\ref{Sec:host}, we model the host galaxy and derive the SMBH mass. In Section~\ref{sec:data}, we describe the photometric and spectroscopic data reduction. In Section~\ref{sec:photometry}, we analyze the photometric properties, including thermal properties and timescales. In Section~\ref{sec:spectroscopy}, we analyze the spectra and determine the evolution of spectral lines. In Section~\ref{sec:disc}, we discuss the interpretation. In Section~\ref{sec:thirdflare}, we present an additional explanation of the missing third flare. Finally, in Section~\ref{sec:conc}, we summarize our results.

We assume a flat $\Lambda$CDM cosmology with $H_0=70~\mathrm{km~s^{-1}~Mpc^{-1}}$, $\Omega_M = 0.3$, and $\Omega_\Lambda = 0.7$. At the redshift of TDE\,2022dbl ($z = 0.0284$), this corresponds to a luminosity distance of 124 Mpc. All magnitudes are reported on the AB system. We correct for Galactic extinction using the \citet{fitzpatrick1999correcting} extinction law with $R_V = 3.1$ and $A_V = 0.052$~mag \citep{schlafly2011measuring}. The extinction correction applied to each photometric band is listed in Table~\ref{Tab:Extinctions} in the Appendix. We adopt peak times of MJD 59641.17 $\pm$ 0.23 and 60348.39 $\pm$ 1.05 for the first and second flares, respectively, based on the light-curve fits described in Section~\ref{subsec:riseanddec}.


\section{Host Galaxy Properties}\label{Sec:host}
TDE\,2022dbl is located at $\alpha_{\mathrm{J2000}} =
12^{\mathrm h}20^{\mathrm m}45.01^{\mathrm s}$, $\delta_{\mathrm{J2000}} =
+49^\circ33'04.68''$, at the nucleus of the galaxy SDSS~J122045.04+493304.6 at $z = 0.02840 \pm 0.00001$ \citep{2022TNSAT2022dbl, lin2024unluckiest}, corresponding to a luminosity distance of 124 Mpc.

The host galaxy of TDE\,2022dbl is bright in the optical and UV, and contaminates the observed TDE emission at late times. We therefore model and subtract the host contribution to isolate the transient light. The host has been observed from UV to IR, including \textit{NUV} photometry from the Galaxy Evolution Explorer (GALEX; \citealp{martin2005galaxy}), \textit{ugriz} photometry and spectroscopy from the Sloan Digital Sky Survey (SDSS; \citealp{ahumada202016th}), \textit{JHK} photometry from the Two Micron All-Sky Survey (2MASS; \citealp{skrutskie2006two}), and W1 and W2 photometry from the Wide-field Infrared Survey Explorer (WISE; \citealp{wright2010wide}).

\subsection{Host Galaxy Modeling} \label{sec:hostmodel}
To characterize the host galaxy of TDE\,2022dbl, we modeled its spectral energy distribution (SED) using the \texttt{Prospector} inference framework \citep{leja2017deriving, Johnson2021stellar}. The fit incorporated both photometric and spectroscopic data to constrain the stellar mass, $\log(M/M_\odot)$; stellar population age, $t_{\rm age}$; metallicity, $\log(Z/Z_\odot)$; dust attenuation, \texttt{dust2}; and the characteristic star-formation decline timescale, $\tau$. We used Kron fluxes measured in elliptical apertures for the GALEX-\textit{NUV} and 2MASS bands, Petrosian magnitudes for the SDSS bands, and 8.5\arcsec-aperture photometry for the WISE bands, chosen to enclose as much host-galaxy light as possible.

We adopted a parametric star formation history (SFH) with dust emission, implemented through the Flexible Stellar Population Synthesis (FSPS; \citealp{conroy2009propagation, conroy2010propagation}) code with default priors. The redshift was fixed, and the SDSS spectrum was scaled to account for the fiber aperture correction. The input photometry is summarized in Table \ref{Tab:Prospector input}, and the corresponding filter transmission curves were obtained from the SVO Filter Profile Service \citep{rodrigo2012svo, Rodrigo2020svo}.

The parameter estimation was performed using \texttt{dynesty}, a nested-sampling algorithm. The best-fit parameters are $\log(M/M_\odot) = 10.22^{+0.01}_{-0.01}$, $\log(Z/Z_\odot) = -0.64^{+0.01}_{-0.01}$, $\texttt{dust2} = 0.0011^{+0.0018}_{-0.0008}$\,mag, $t_{\rm age} = 6.74^{+0.23}_{-0.30}$\,Gyr, and $\tau = 0.54^{+0.05}_{-0.09}$\,Gyr. Using the surviving stellar-mass fraction from \texttt{Prospector}, \texttt{mfrac}=0.60, we compute the present-day host-galaxy stellar mass as $M_{\rm gal}=M\times\texttt{mfrac} = 10.00^{+0.01}_{-0.01}$.

\begin{figure*}
  \centering
  \includegraphics[width=\linewidth]{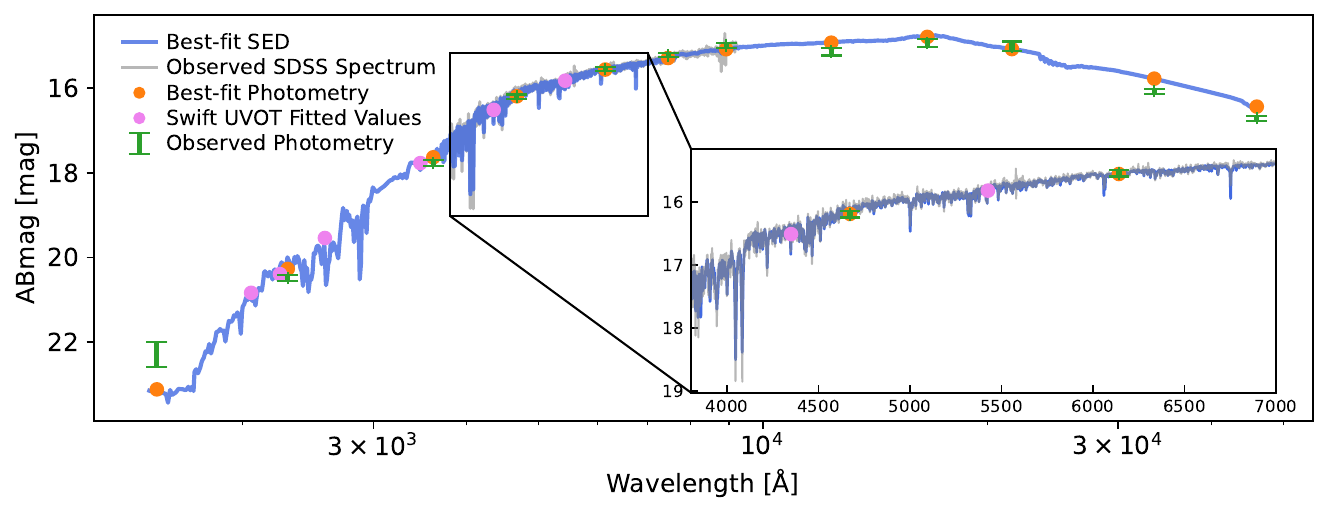}
  \caption{Comparison of the best-fit SED (blue) with the observed SDSS spectrum (gray). The photometric data used in the fit are shown with green error bars, whereas the corresponding best-fit photometric values are shown in orange. Violet points indicate the predicted host-galaxy fluxes in the \textit{Swift}-UVOT bands from the fitted model. The inset zooms to show that the model reproduces the observed continuum and major host-galaxy absorption features.
}
  \label{fig:host_spectrum}
\end{figure*}

These values are consistent with an early-type galaxy. The shallow Balmer absorption lines and the absence of nebular emission lines indicate little or no ongoing star formation. The galaxy likely contains little dust, because its red color is primarily driven by an aging stellar population rather than by extinction. The derived stellar mass is consistent with the SDSS DR16 Granada FSPS VAC value, $\log(M/M_\odot) = 10.2470 \pm 0.0060$, from the FSPSGranEarlyNoDust fit \citep{ahn2014tenth, ahumada202016th}.

The predicted spectrum reproduces the observations well (Figure~\ref{fig:host_spectrum}), and the posterior distributions indicate that the model parameters are tightly constrained (Figure~\ref{fig:corner}). The effective sample size is 10,293, with a final Bayesian evidence of $\ln Z = 62679.01\pm0.16$, indicating a well-sampled posterior and a precise evidence estimate.

\subsection{Velocity Dispersion Estimation}
The SDSS spectrograph has an instrumental resolution of $R \sim 2000$, corresponding to a velocity dispersion resolution of $\approx$ 64 km $\text{s}^{-1}$. The SDSS DR16 pipeline reports a stellar velocity dispersion of $60.0 \pm 5.1$ km $\text{s}^{-1}$ for this galaxy, which is comparable to the instrumental resolution. To obtain a more robust measurement and thereby improve the black hole mass estimate, we obtained medium-resolution spectroscopy with Keck/ESI on 2023 April 15, during the post-flare phase when the host dominated the optical emission (+408 days after the first flare peak). Observations were taken using a 0.75\arcsec\ slit, yielding a spectral resolution of $R = 5400$ (velocity resolution $\sigma_v \approx$ 24 km $\text{s}^{-1}$), more than sufficient to robustly resolve velocity dispersions at this scale.

We reduced the spectroscopic data following \citet{wevers2019black}. The MAuna Kea Echelle Extraction(\texttt{MAKEE}) software package was used for standard processing steps, including bias subtraction, flat-field correction, wavelength calibration, and spectrum extraction. After an initial extraction based on the trace, we refined the extraction by selecting the host galaxy bulge as the source region and defining appropriate background regions. We adopted an extraction radius of 14.7 pixels (2.26\arcsec), corresponding to the host-galaxy effective radius of 1.29~kpc \citep{simard2011catalog}, to capture the bulge emission. The spectrum was rebinned onto a logarithmic wavelength scale with a dispersion of 11.5 km s$^{-1}$ pixel$^{-1}$ using the \texttt{combine} command.

\begin{figure}[h!]
  \centering     
  \includegraphics[width=\columnwidth]{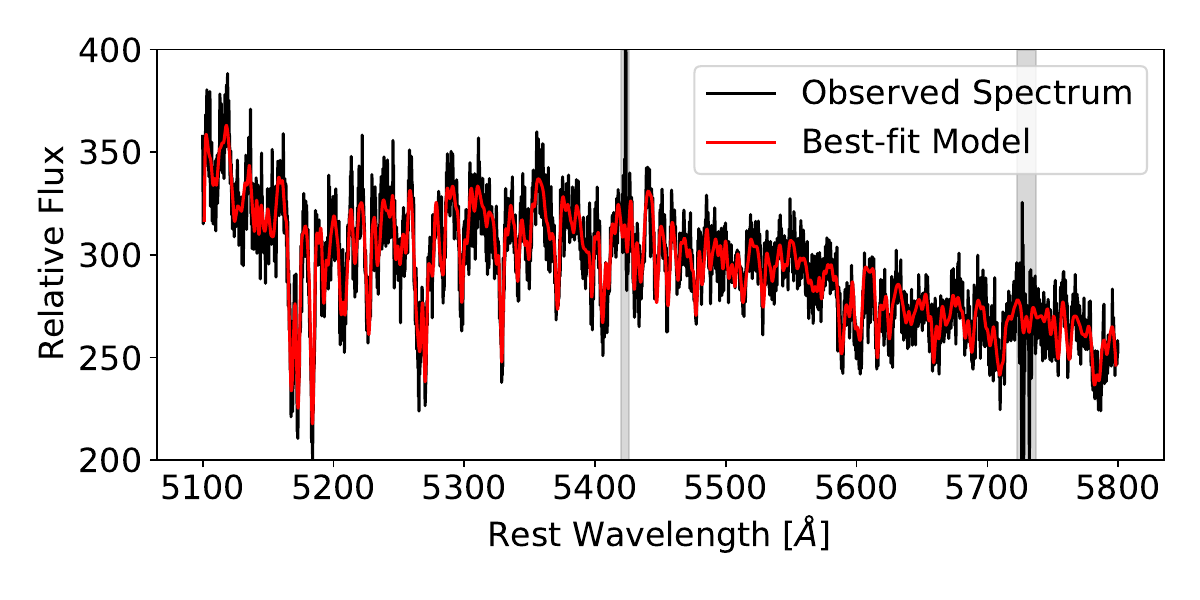}
  \caption{Stellar velocity dispersion fit using \texttt{pPXF}. The black line shows the observed rest-frame ESI spectrum of the host galaxy's bulge, while the red line represents the best-fit \texttt{pPXF} model using ELODIE stellar templates convolved to the ESI resolution. The gray regions mark wavelengths of bright night sky lines masked from the fit.}
  \label{fig:ppxffig}
\end{figure}

We measured the stellar velocity dispersion of the bulge with the Penalized Pixel Fitting (\texttt{pPXF}) package \citep{cappellari2017improving}. We adopted 1959 high-resolution spectra from the ELODIE library as templates \citep{prugniel2001database, prugniel2007new} and convolved them to match the ESI instrumental resolution. We used the default fourth-order polynomial for continuum fitting. To mitigate contamination from strong sky emission and interstellar absorption, we masked regions at 5420--5426~\AA, 5724--5736~\AA, and 5885--5900~\AA\ in the rest frame. The resulting fit is shown in Figure \ref{fig:ppxffig}. We obtain a velocity dispersion of $53.3 \pm 2.1$ km s$^{-1}$. Varying the continuum polynomial degree produced consistent results, confirming the robustness of the measurement. It is also consistent with the SDSS value of $60.0 \pm 5.1$ km s$^{-1}$, supporting a low-dispersion host galaxy.

\subsection{SMBH Mass} \label{sec:SMBHmass}
We estimated the SMBH mass using the $M$--$\sigma$ relations from \citet{greene2020intermediate, mcconnell2013revisiting, kormendy2013coevolution} and the $M_{\rm BH}$--$M_{\rm gal}$ relations from \citet{yao2023tidal, greene2020intermediate}. The results are summarized in Table~\ref{tab:SMBHmass}.

\begin{deluxetable}{lcc}
\tablecaption{The top panel shows the SMBH mass estimates derived from the $M$–$\sigma$ relation for early-type galaxies, with emphasis on systems with small velocity dispersions. We consider this relation the most reliable and use it for subsequent calculations and analysis. The middle panel presents additional SMBH mass estimates using other $M$–$\sigma$ relations available in the literature. The bottom panel shows the mass estimates based on the $M_{\rm BH}$–$M_{\rm gal}$ relations. \label{tab:SMBHmass}}
\tablehead{
\colhead{} & \colhead{Galaxy Type} & \colhead{$\log_{10} \left( \frac{M_{BH}}{M_\odot} \right)$}
}
\startdata
\citet{greene2020intermediate} & Early Type & $6.01 \pm 0.45$ \\
\hline
\citet{greene2020intermediate} & All Galaxies & $5.70 \pm 0.57$ \\
\citet{mcconnell2013revisiting} & Early Type & $5.40 \pm 0.41$ \\
\citet{mcconnell2013revisiting} & All Galaxies & $5.08 \pm 0.44$ \\
\citet{kormendy2013coevolution} & & $5.98 \pm 0.35$ \\
\hline
\citet{yao2023tidal} &  & $6.80 \pm 0.22$ \\
\citet{greene2020intermediate} & Early Type & $7.26 \pm 0.66$ \\
\citet{greene2020intermediate} & All Galaxies & $6.90 \pm 0.80$ \\
\enddata
\end{deluxetable}
These methods produce a range of SMBH mass estimates due to differences in sample selection and calibration. We adopt the early-type $M$–$\sigma$ relation from \citet{greene2020intermediate} because their sample explicitly targets low-mass SMBHs in galaxies comparable to our host. In contrast, the \citet{mcconnell2013revisiting} and \citet{kormendy2013coevolution} samples do not probe velocity dispersions as low as our measured value. We also favor $M$–$\sigma$ relations over $M_{\mathrm{BH}}$–$M_{\rm gal}$ relations, as velocity dispersion shows a tighter empirical correlation with SMBH mass. Among the $M_{\mathrm{BH}}$–$M_{\rm gal}$ estimates, the relatively small scatter reported by \citet{yao2023tidal} may reflect their limited sample size (19 objects).

Using the \citet{greene2020intermediate} relation, we derive $\log(M_{\mathrm{BH}}/M_\odot) = 6.01 \pm 0.45$. This is consistent with the $\log(M_{\mathrm{BH}}/M_\odot)\approx 6.4$ values reported by \citet{lin2024unluckiest} and \citet{hinkle2024double} based on the SDSS velocity dispersion. It is also lower than the \verb+MOSFiT+ \citep{guillochon2018mosfit} estimate of $\log(M_{\mathrm{BH}}/M_\odot) \sim 6.65 \pm 0.20$ from \citet{makrygianni2025double}. \citet{angus2026can} used \texttt{MOSFiT} with priors chosen to specifically probe the pTDE scenario, obtaining $\log(M_{\mathrm{BH}}/M_\odot) = 6.00^{+0.02}_{-0.05}$ and $6.77^{+0.03}_{-0.14}$ for the first and second flares, respectively. Our value agrees closely with their first-flare estimate, while the higher second-flare value may indicate that the second flare does not follow a simple fallback relation. We note that $M_{\rm BH}$ inferred from \texttt{MOSFiT} is subject to degeneracies with $\beta$ and the mass of the disrupted star, and to systematic underestimation for a poorly sampled light curve \citep{mockler2019weighing, angus2026can}.

Our estimated SMBH mass is consistent with several model-dependent estimates. These include \verb+TDEmass+ \citep{ryu2020measuring} estimates based on peak luminosity and color of $\log(M_{\mathrm{BH}}/M_\odot)=6.36\pm0.40$ and $6.04\pm0.41$ for the first and second flares, respectively \citep{angus2026can}; a cooling-envelope fit with \verb+Redback+ \citep{sarin2024redback} of $5.41^{+0.70}_{-0.48}$ \citep{angus2026can}; and a thin-disk plateau-luminosity estimate of $5.68^{+0.60}_{-0.44}$ \citep{mummery2024fundamental, mummery2025optical}. All are consistent with our measurement within their uncertainties.

\section{OBSERVATIONS}\label{sec:data}
\subsection{Swift-UVOT photometry}\label{SubSec:dataUVOT}
The Ultra-Violet/Optical Telescope (UVOT; \citealt{roming2005swift}) onboard the Neil Gehrels Swift Observatory \citep{gehrels2004swift} began observing TDE\,2022dbl on 2022 February 27 (PIs: Hinkle, Holoien, Margutti, Lin, Shang). We analyzed 131 epochs spanning MJD 59637 to 61079. Most epochs include all six UVOT filters: $v$ (5425.3~\AA), $b$ (4349.6~\AA), $u$ (3467.0~\AA), uvw1 (2580.7~\AA), uvm2 (2246.4~\AA), and uvw2 (2054.6~\AA; pivot wavelengths). These data are affected by instrumental systematics such as image trailing in some epochs, potentially impacting the accuracy of the standard \swift-UVOT reduction pipeline.

We retrieved the data from HEASARC and re-reduced them using \texttt{HEASoft v6.32} and the most recent \texttt{CALDB}. Observations with evidence of lost star tracking were removed, and \texttt{uvotimsum} was used to combine image extensions within each epoch. We used a 10\arcsec\ source aperture for all filters and 40\arcsec--50\arcsec\ source-free background regions. Photometry was extracted using \texttt{uvotsource}. To ensure a signal-to-noise ratio (S/N) $>10$, we merged low S/N observations by summing up their extensions and then performing \texttt{uvotsource}.

Because the UVOT pipeline PSF correction is calibrated for 5\arcsec\ source apertures \citep{poole2008photometric}, larger apertures can lead to underestimated uncertainties. To account for this, we selected four nearby non-variable sources and measured their photometry using the same procedure. For each band, we kept observations with S/N~$>$~5 and determined a correction factor $\alpha$ such that the reduced $\chi^2$ of the light curves for these reference sources satisfied $\alpha \times \chi^2_\nu = 1$. The mean $\alpha$ values are reported in Table~\ref{tab:alphacorrection}, and were used to rescale the photometric uncertainties.

Due to \emph{Swift}'s extended PSF, some host-galaxy flux might fall outside the 10\arcsec\ aperture. To account for flux redistributed into the PSF wings, we constructed radial profiles using concentric circular apertures from 0\arcsec\ to 50\arcsec\ in 2\arcsec\ steps with SAOImage DS9 \citep{joye2003new}. The profiles were fit with a \texttt{Moffat1D} model plus a constant background \citep{moffat1969theoretical}. From the best-fit model, we determined the fraction of flux enclosed within 10\arcsec. This was applied to the \verb+Prospector+ host model to obtain the PSF-corrected host galaxy \emph{Swift}-UVOT photometry. The results are listed in Table~\ref{Tab:PSFcorrection}.

After correcting for Galactic extinction, we subtract the PSF-corrected host flux from the \emph{Swift}-UVOT photometry. For non-detections (S/N~$\leq$~3), we report $3\sigma$ upper limits. The final photometry is listed in Table~\ref{Tab:Swift UVOT}. The full multi-band light curve is shown in Figure~\ref{fig:lightcurve}, with individual filter light curves presented in Figure~\ref{fig:Swift_subplots}.

\subsection{Optical photometry}\label{SubSec:opticalphot}
We obtained $g$, $r$, and $i$-band photometry from the Zwicky Transient Facility (ZTF; \citealt{Bellm2019, Graham2019, Dekany2020}) between March 2018 and March 2026 from the ZTF Forced Photometry Service \citep{masci2023new}. The data were reduced following the methods discussed in \citet{Masci2019ZTF}. We removed poor-quality epochs with seeing cut (\texttt{sciinpseeing}) $< 4$, the spatial noise-sigma per pixel (\texttt{scisigpix}) $>25$, or poor photometric calibration (\texttt{infobitssci}) $\geq 33554432$. Baseline subtraction and Galactic extinction corrections were then applied. The data were binned using a maximum bin width of $\sim 30$ days, or until the bin reached S/N $\geq 3$. 

We also obtained ATLAS $c$- and $o$-band forced photometry from December 2015 to March 2026 \citep{2018PASP..130f4505T,2020PASP..132h5002S} through the ATLAS forced photometry server \citep{Shingles2021release}. The photometry was measured on difference images. We removed outliers and corrected the photometry for Galactic extinction. We first combined the measurements into 5-day bins using inverse-variance weighted averages of the fluxes to reduce scatter. The ATLAS data were then further binned using the same binning and detection-threshold procedure as for the ZTF data.

\begin{figure*}
  \centering
  \includegraphics[width=1.98\columnwidth]{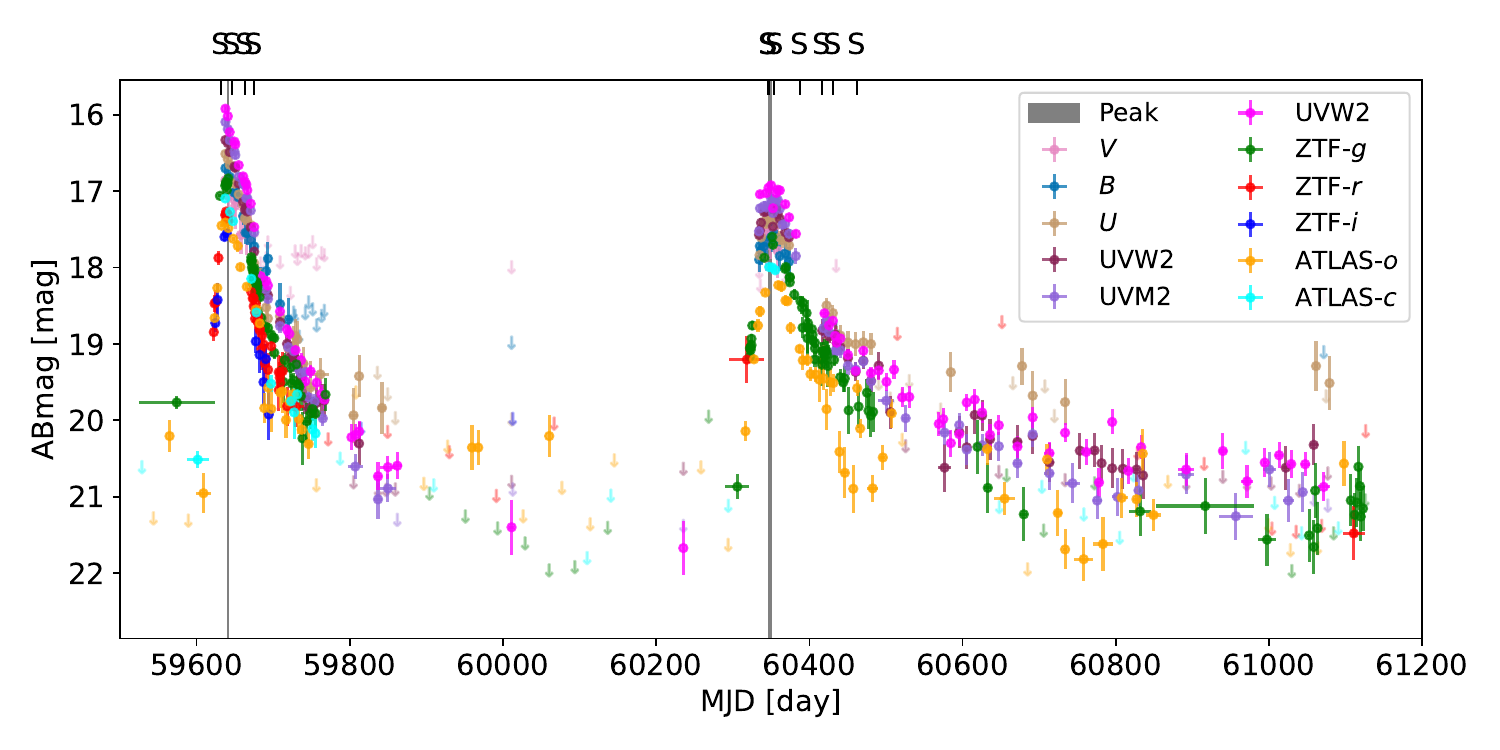}
  \caption{UV and optical light curves of TDE\,2022dbl from \swift-UVOT, ZTF, and ATLAS, corrected for Galactic extinction and host-galaxy emission. Downward arrows indicate $3\sigma$ upper limits. Dark gray vertical bands mark the peak times of the first and second flares from \citet{makrygianni2025double}. Top ticks labeled ``S'' indicate the epochs of optical spectroscopic observations.}    
  \label{fig:lightcurve}
\end{figure*}

The final optical photometry between MJD 59500 and 60500 is shown in Figure \ref{fig:lightcurve}.

\subsection{Swift-XRT}\label{SubSec:dataXRT}
The \emph{Swift} X-Ray Telescope (XRT; \citealt{Burrows05}) provided good coverage of both flares of AT2022dbl. We downloaded the XRT data from HEASARC and followed standard data reduction practices with \texttt{HEASoft v6.36} and the most recent \texttt{CALDB} (e.g., \citealt{Margutti13}).

No X-ray emission is detected during the first flare. A weak detection is found at $\sim372$ days after the first peak (corresponding to ObsIDs 00015026043 and 00015026044), with a 0.3--10 keV count rate of $3.3^{+1.8}_{-1.3}\times10^{-3}$ counts s$^{-1}$. No significant X-ray emission is detected during the second flare. These results are consistent with those reported by \citet{hinkle2024double}.

We extracted an X-ray spectrum during the weak-detection interval (368.5--371.5 days after the first peak; exposure time of 2.3\,ks ) and fit it with an absorbed blackbody model. Only six photons remained after the quality cuts, so the spectral parameters are poorly constrained. We fit the spectrum with the \verb+tbabs*bbody+ model, fixing the Galactic absorption to $N_{\rm H}=1.94\times10^{20}$ cm$^{-2}$ \citep{HI4PI}. The best-fitting temperature is $kT=0.024\pm0.010$ keV, indicating a very soft thermal spectrum. Fixing the temperature to this nominal best-fitting value, we infer an observed 0.3--10 keV flux of $F_{\rm X,obs}=1.02^{+0.69}_{-0.32}\times10^{-13}$ erg cm$^{-2}$ s$^{-1}$, corresponding to an observed luminosity of $L_{\rm X,obs}\approx1.9\times10^{41}$ erg s$^{-1}$. After correcting for absorption, the inferred unabsorbed luminosity is $L_{\rm X,unabs}\approx3.0\times10^{41}$ erg s$^{-1}$.

We also include new \swift-XRT observations obtained after the second flare. None of the individual epochs shows a clear detection. Stacking all epochs between MJD 60706 and 61079, corresponding to 358--731 days after the second peak, yields a $3\sigma$ upper limit of $3.06\times10^{-4}$ counts s$^{-1}$ in the 0.3--10 keV band, which corresponds to $F_{\rm X,obs}<9.4\times10^{-14}$ erg s$^{-1}$ cm$^{-2}$ ($L_{\rm X,unabs}<2.7\times10^{40}$ erg s$^{-1}$ for the spectral parameters of the detection).  

\subsection{Optical spectroscopy}\label{SubSec:spectra}
Eleven spectra were obtained to track the spectroscopic evolution of TDE\,2022dbl, comprising 1 spectrum from Faulkes Telescope North (FTN)/FLOYDS (acquired from TNS; \citealt{2022TNSCR.504....1A}), 1 from Lowell Discovery Telescope (LDT)/DeVeny, 6 from the Shane 3\,m telescope/Kast \citep{millerstone93}, and 3 from the Low-Resolution Imaging Spectrometer on the Keck-I 10~m telescope (LRIS; \citealt{oke1995keck}). The log of optical spectroscopic observations of TDE\,2022dbl is provided in Table \ref{Tab:speclog}.

The DeVeny observations were taken with the 2$\arcsec$-wide slit using the 300 g/mm DV2 grating with a central wavelength of 5800\AA. The spectrum was reduced using \texttt{PypeIt} \citep{Prochaska2020_pypeit,Prochaska_pypeit_zenodo} and standard optical spectroscopic reduction techniques, including bias subtraction, flat-fielding, flux calibration, coaddition, and telluric correction.

The Kast observations used a 600/4310 grism on the blue side, D57 dichroic beamsplitter, and the 600/7500 grating on the red side, along with a 2$\arcsec$-wide slit, resulting in $\sim$5\,\AA\ resolution over the range 3500--8750~\AA. The LRIS observations used the 600/4000 grism on the blue side, the d560 dichroic beamsplitter, and the 400/8500 grating on the red side to cover the full optical range of 3150--10300\,\AA. With a 1$\arcsec$-wide slit, this resulted in resolutions of $\sim$4 and 6~\AA\ on the blue and red sides, respectively. The data reduction followed standard procedures, as outlined by \citet{silverman2012}. Fixed spectral extraction apertures around the galaxy nucleus ($\pm2\arcsec$ for Kast, $\pm1\arcsec$ for LRIS) were used for each instrument to minimize variations in host galaxy contamination.

To isolate the transient emission, we corrected for Galactic extinction and subtracted the host contribution. We fit the archival SDSS spectrum of the host galaxy using \texttt{pPXF} \citep{cappellari2017improving} and adopted the best-fit model as the unscaled host template. For each spectrum, we fit a linear combination of the interpolated host model and a blackbody continuum to featureless regions, excluding prominent emission lines. The scaled host component from this fit was then subtracted from the observed data. The fitting regions are $<$3900~\AA, 4300--4400~\AA, 5150--6300~\AA, and 7100--7300~\AA. For the $-0.89$~day spectrum relative to the second peak, the first region is restricted to 3800--3900~\AA\ due to the steep slope at the blue end.

\begin{figure*}[t]
  \centering
  \includegraphics[width=0.92\linewidth,page=1] {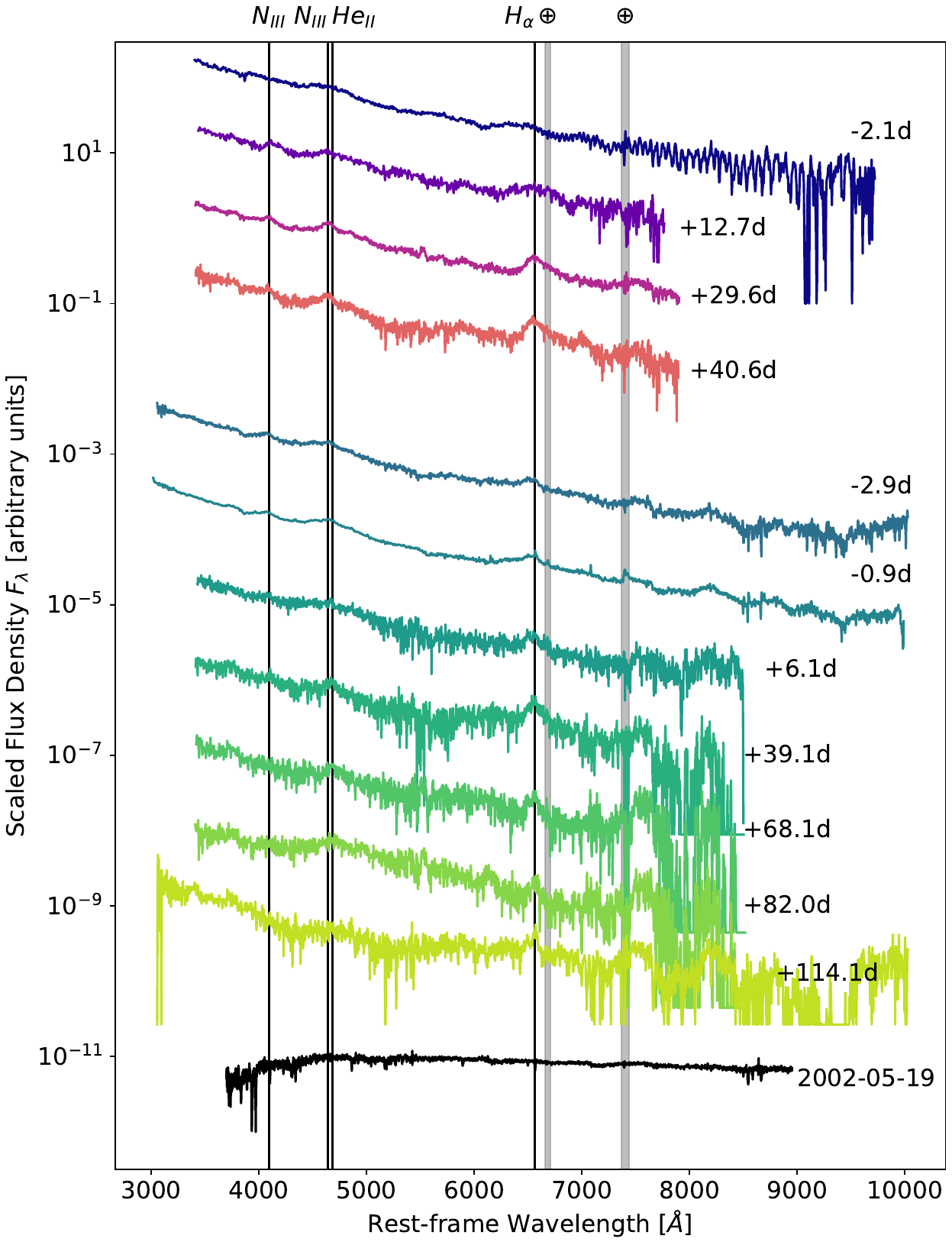}
  \caption{Optical spectra of TDE\,2022dbl obtained with FTN/FLOYDS, LDT, Kast/Shane, and Keck/LRIS. The colored transient spectra are shown in chronological order, with the first four from the top corresponding to the first flare and the following seven corresponding to the second flare. The bottom black spectrum shows the SDSS archival host-galaxy spectrum. The spectra have been multiplied by arbitrary scaling factors for visualization while preserving the relative spectral shapes.}
  \label{fig:spectratotal}
\end{figure*}

We scaled the TDE-only spectra to match contemporaneous ZTF $g$-band photometry, ensuring consistent flux calibration across epochs. Figure~\ref{fig:spectratotal} shows the resulting TDE-only emission spectra.

\section{Photometric Properties} \label{sec:photometry}
TDE\,2022dbl exhibits two flares peaking at MJD $59641.17\pm0.23$ and $60348.4\pm1.0$ (Figure~\ref{fig:lightcurve}). The emission is initially blue ($g-r$ mag $\sim-0.48$ mag), with the \swift-uvw2 band brightest, and evolves to redder colors over time. The first flare is brighter and more sharply peaked than the second. 

At late times after the first peak, the source is only seen in \swift-uvw2, with an apparent magnitude of 21.67 ($M_{\text{uvw2}}=-13.77$) at 594.3 days after peak, indicating a persistent late-time UV emission. Because this detection occurs relatively early and is brighter than typical late-time plateau values reported for other TDEs \citep{van2019late, mummery2024fundamental, mummery2025optical}, it may represent emission still declining toward a plateau, possibly associated with the formation of an accretion disk, rather than a fully settled late-time state.

In contrast to the first flare, the second flare remains detected at comparable phases in multiple UV/optical bands. During the late-time UV-excess phase, the \swift-uvw2 luminosity ratio between the second and first flares increases from $2.44\pm0.88$ at $\sim370$ days after peak to $3.2\pm1.2$ at $\sim600$ days after peak. The bolometric comparison is less well constrained, with $L_{\rm bol,2}/L_{\rm bol,1}$ increasing from $1.12^{+0.48}_{-0.38}$ to $4.5^{+8.8}_{-3.1}$ over the same interval. This suggests a more persistent UV/optical-emitting component after the second flare. One possible interpretation is that repeated disruptions provide additional material to the accretion flow, producing a more luminous and longer-lived disk component. 

\subsection{Optical/UV Spectral Energy Distribution}\label{sec:blackbody}

To characterize the thermal properties of TDE\,2022dbl, we modeled the SED with a blackbody and determined temperature ($T_{\rm bb}$) and radius ($R_{\rm bb}$) at each \textit{Swift} epoch. Photometric data with ${\rm S/N}>3$ from \textit{Swift}-UVOT, ZTF, and ATLAS were included in the fits. A Gaussian Process interpolation with a \verb+Matérn+ kernel \citep{Rasmussen2006}, implemented with the \texttt{scikit-learn} package \citep{Pedregosa2011skitlearn}, was used to estimate fluxes between epochs.

\begin{figure}[h!]
  \centering
  \includegraphics[width=\columnwidth]{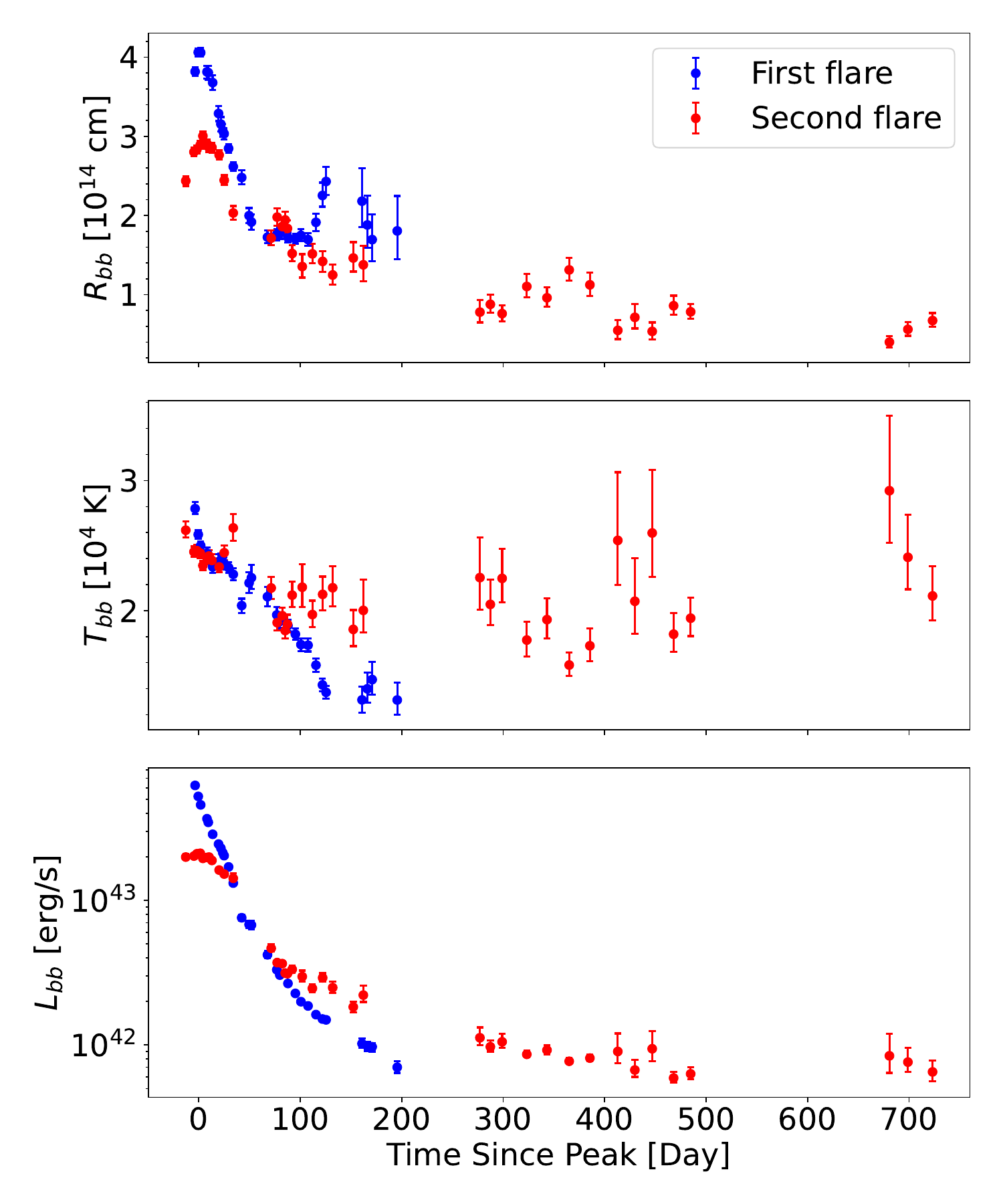}
  \caption{Blackbody evolution of TDE\,2022dbl. The top, middle, and bottom panels show the blackbody radius $R_{\rm bb}$, temperature $T_{\rm bb}$, and luminosity $L_{\rm bb}$, respectively. Blue and red points correspond to the first and second flares, respectively.}
  \label{fig:blackbody}
\end{figure}

The best-fit parameters were obtained using the nested sampling Monte Carlo algorithm \texttt{MLFriends} \citep{buchner2016statistical, buchner2019collaborative}, as implemented in the \texttt{UltraNest} package\footnote{\url{https://johannesbuchner.github.io/UltraNest/}} \citep{buchner2021ultranest}. Temperatures are reported in the rest frame. The resulting evolution of $R_{\rm bb}$ and $T_{\rm bb}$ is shown in Figure~\ref{fig:blackbody} and summarized in Table~\ref{Tab: blackbody}.

In the top panel of Figure~\ref{fig:blackbody}, $R_{\rm bb}$ rises briefly before declining, reaching peak values of $(4.061 \pm 0.098)\times10^{14}$ cm and $(3.00 \pm 0.11)\times10^{14}$ cm for the first and second flares at $\sim-0.4$ and $\sim4.1$ days, respectively. These values are consistent within the uncertainties with those reported by \citet{lin2024unluckiest}, but are more tightly constrained. The middle panel shows the corresponding temperature evolution. Near peak, the first and second flares reach maximum rest-frame temperatures of $27840 \pm 920$ K and $26200 \pm 1200$ K, respectively. Both flares cool after peak: the first declines to $\sim13000$ K, while the second initially follows a similar trend but later remains near $\sim20000$ K. At late times, the second flare shows a possible temperature upturn while $R_{\rm bb}$ continues to decrease. Although the uncertainties are large, this behavior may indicate that the effective photosphere is receding. The bottom panel shows the corresponding bolometric luminosity evolution, calculated from the best-fit $T_{\rm bb}$ and $R_{\rm bb}$. The first and second flares reach peak luminosities of $L_{\rm bb} = (6.24\pm0.24)\times10^{43}$ erg s$^{-1}$ and $(2.117\pm0.067)\times10^{43}$ erg s$^{-1}$, respectively. These values lie at the lower-luminosity end of the TDE population and agree with those reported by \citet{makrygianni2025double}. The luminosities of the two flares remain within a factor of three of each other, with the first flare peaking nearly three times brighter and declining more rapidly. The second flare remains detected at late times at $L_{\rm bb}\sim10^{42}$ erg s$^{-1}$, suggestive of plateau-like emission.

The rise of both flares is assumed to have a constant color, scaled to ZTF $g$-band data. Before the first detection, we extrapolate the luminosity using Equation~\ref{eq:luminosity}. Based on the blackbody parameters, the integrated bolometric energies out to 195.6 days after peak are $(2.030\pm0.031)\times10^{50}$~erg and $(1.508\pm0.037)\times10^{50}$~erg for the first and second flares, respectively. These values are 1.3 and 1.9 times those reported by \citet{makrygianni2025double}, respectively, likely because our integration extends over a longer time range.

Assuming the typical radiative efficiency $\eta = 0.1$ \citep{xie2012radiative}, the inferred accreted masses are $(1.136 \pm 0.018)\times 10^{-3} M_\odot$ and $(0.844 \pm 0.021)\times 10^{-3} M_\odot$ for the first and second flares, respectively. These inferred masses are much smaller than the mass of even a low-mass star, pointing to the missing-energy problem \citep{piran2015circularizationvsaccretion, Lu_2018}. However, this discrepancy may be less severe in the context of a pTDE interpretation. If we instead integrate the second flare to 723 days after peak, including the late-time plateau-like emission, the total radiated energy increases to $(1.942\pm0.060)\times10^{50}$~erg. If the plateau persists, it corresponds to an average accretion rate of $\sim1.7\times10^{-4}M_{\odot}{\rm yr}^{-1}$ for $\eta=0.1$. Continued late-time emission could therefore partly alleviate the missing energy problem, although the total radiated energy remains well below the rest-mass energy of a substantial fraction of a star.

\begin{figure}[h!]
  \centering
  \includegraphics[width=\columnwidth]{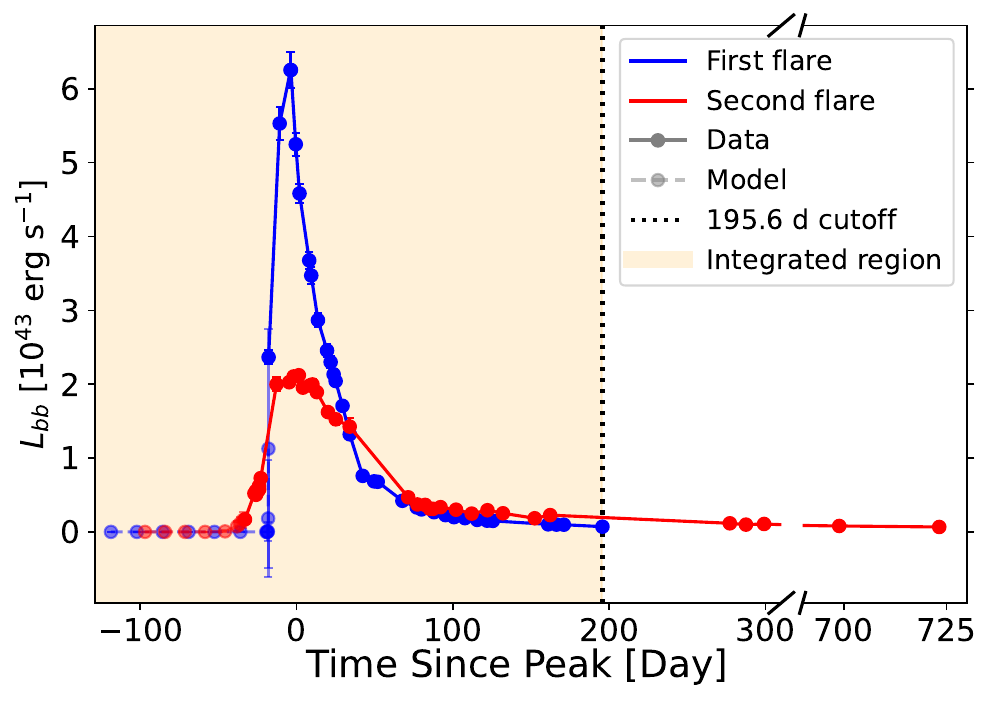}
  \caption{Bolometric luminosity evolution of the first (blue) and second (red) flares of TDE\,2022dbl. For the rise, we assume constant color and estimate the luminosity using the ZTF $g$-band photometry. Prior to the first well-detected rising data point, we extrapolate the luminosity using the power-law model from Equation~\ref{eq:luminosity}, shown by the dashed, faded lines. The shaded region marks the interval used to compute the bolometric energy, extending to 195.6 days after peak, corresponding to the last well-constrained observation of the first flare. A broken $x$-axis is used to emphasize the near-peak evolution while still showing the late-time second-flare detections; little luminosity evolution is observed between $\sim300$ and $\sim700$ days after the second peak.}
  \label{fig:bolometric}
\end{figure}

\subsection{Rise and Decline Timescales}\label{subsec:riseanddec}
The rise and decline timescales of a TDE light curve provide important constraints on the underlying emission mechanisms, including the fallback of stellar debris, radiative diffusion through the reprocessing layer, and the efficiency of energy release.

We followed the approach of \citet{van2021seventeen, yao2023tidal} to model the rise and decline in the ZTF-$g$ band using
\begin{equation}
L_\nu(t) = A_\nu \times
\begin{cases}
0 & t \leq t_{\text{fl}} \\
\left( \dfrac{t - t_{\text{fl}}}{t_{\text{peak}} - t_{\text{fl}}} \right)^n & t_{\text{fl}} < t \leq t_{\text{peak}} \\
\exp\left( -\dfrac{t - t_{\text{peak}}}{\tau_{\text{decay}}} \right) & t > t_{\text{peak}}
\end{cases}
\label{eq:luminosity}
\end{equation}where $A_\nu$ is the luminosity normalization, $t_{\mathrm{fl}}$ is the first-light epoch, $t_{\mathrm{peak}}$ is the peak time, and $\tau_{\mathrm{decay}}$ is the decay timescale. From the fit, we obtain peak times of MJD $59641.17 \pm 0.23$ and $60348.39 \pm 1.05$ for the first and second flares, respectively, consistent with the values reported by \citet{makrygianni2025double}.

We define the characteristic timescales $t_{1/2,\mathrm{rise}}$ and $t_{1/2,\mathrm{decline}}$ as the intervals from rising half-maximum to peak and from peak to declining half-maximum, respectively. From the best-fit model, we measured $t_{1/2,\mathrm{rise}} = 17.50 \pm 0.28$ days and $16.11 \pm 1.23$ days, and $t_{1/2,\mathrm{decline}} = 21.62 \pm 0.21$ days and $33.79 \pm 0.70$ days for the first an second flare.

\begin{figure}
  \centering
  \includegraphics[width=\columnwidth]{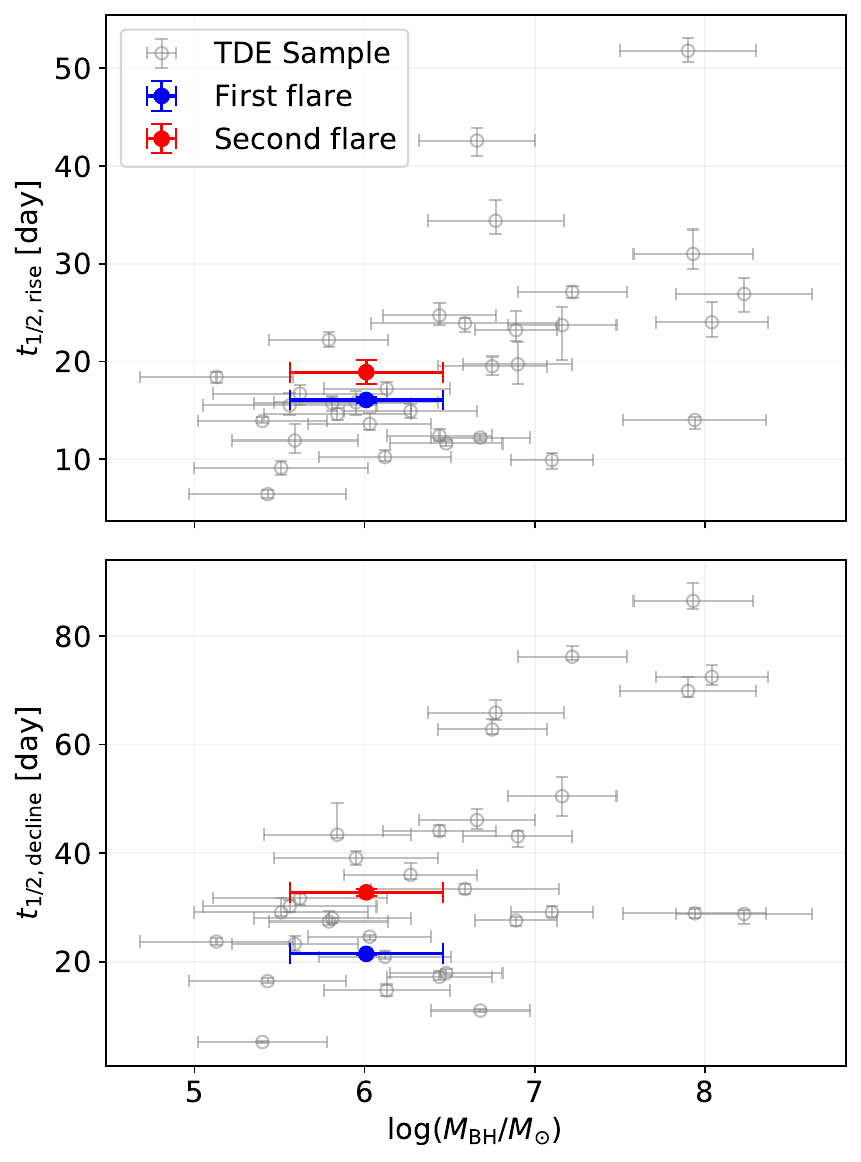}
  \caption{Comparison of the rise (upper) and decline (lower) half-maximum timescales with SMBH mass for the TDE sample from \citet{yao2023tidal}. The two flares of TDE\,2022dbl (blue and red points) are overplotted with their associated uncertainties in both $t_{1/2}$ and $M_{\mathrm{BH}}$. Both flares fall well within the observed distribution of optical TDEs.}
  \label{fig:riseanddecline}
\end{figure}

\citet{yao2023tidal} presented a detailed study of 33 TDEs, comparing $t_{1/2,\mathrm{rise}}$ and $t_{1/2,\mathrm{decline}}$ with inferred SMBH masses. We added the two flares of TDE\,2022dbl to this comparison in Figure~\ref{fig:riseanddecline}. Both flares lie within the overall distribution of observed TDEs, and the second flare shows a notably longer half-decline timescale than the first. These empirical differences are explored further in Section \ref{sec:disc}.

\section{Spectroscopic Properties}\label{sec:spectroscopy}
Spectroscopically, optical/UV TDEs can be classified into several categories based on the presence of hydrogen, helium, and Bowen fluorescence lines (most notably \ion{N}{3} and \ion{O}{3}). A comprehensive spectroscopic classification of TDEs is presented in \citet{charalampopoulos2022detailed}. In this work, we are mainly interested in two broad classes: TDEs that have Bowen fluorescence features in their spectra (Bowen TDEs) and those that do not (non-Bowen TDEs). TDE\,2022dbl belongs to the Bowen TDE class based on the presence of prominent \ion{N}{3} emission. Bowen fluorescence arises from cascades following \ion{He}{2} excitation, and its presence indicates dense, metal-rich gas efficiently reprocessing a copious source of extreme UV radiation. Because these two classes are associated with different physical environments and emission processes, comparing them provides insight into how the environment influences the evolution of other spectral lines.

Figure~\ref{fig:spectratotal} shows that TDE\,2022dbl was spectroscopically monitored from approximately ten days before peak to several months afterward. Both flares exhibit broad H$\alpha$, \ion{N}{3}, and \ion{He}{2} features across all epochs, placing TDE\,2022dbl firmly within the Bowen TDE class. A blue continuum is present throughout the spectra, while the emission-line widths decrease with time. The earliest spectrum displays a strong blue continuum with weak, broad, and blueshifted emission features, likely because an optically thick reprocessing layer obscures the line-forming region \citep{roth2016x, roth2018sets}. 

The two flares differ most strongly near peak, particularly in their luminosities, line widths, and blueshifts, while their spectroscopic properties converge at late times. In the following subsections, we quantify the evolution of the major emission features using parametric line-profile fitting and compare the two flares to study the physical origin of the double-flare behavior.

\subsection{Fits to Emission Features}
\begin{figure*}[t]
  \centering
  \includegraphics[width=\linewidth]{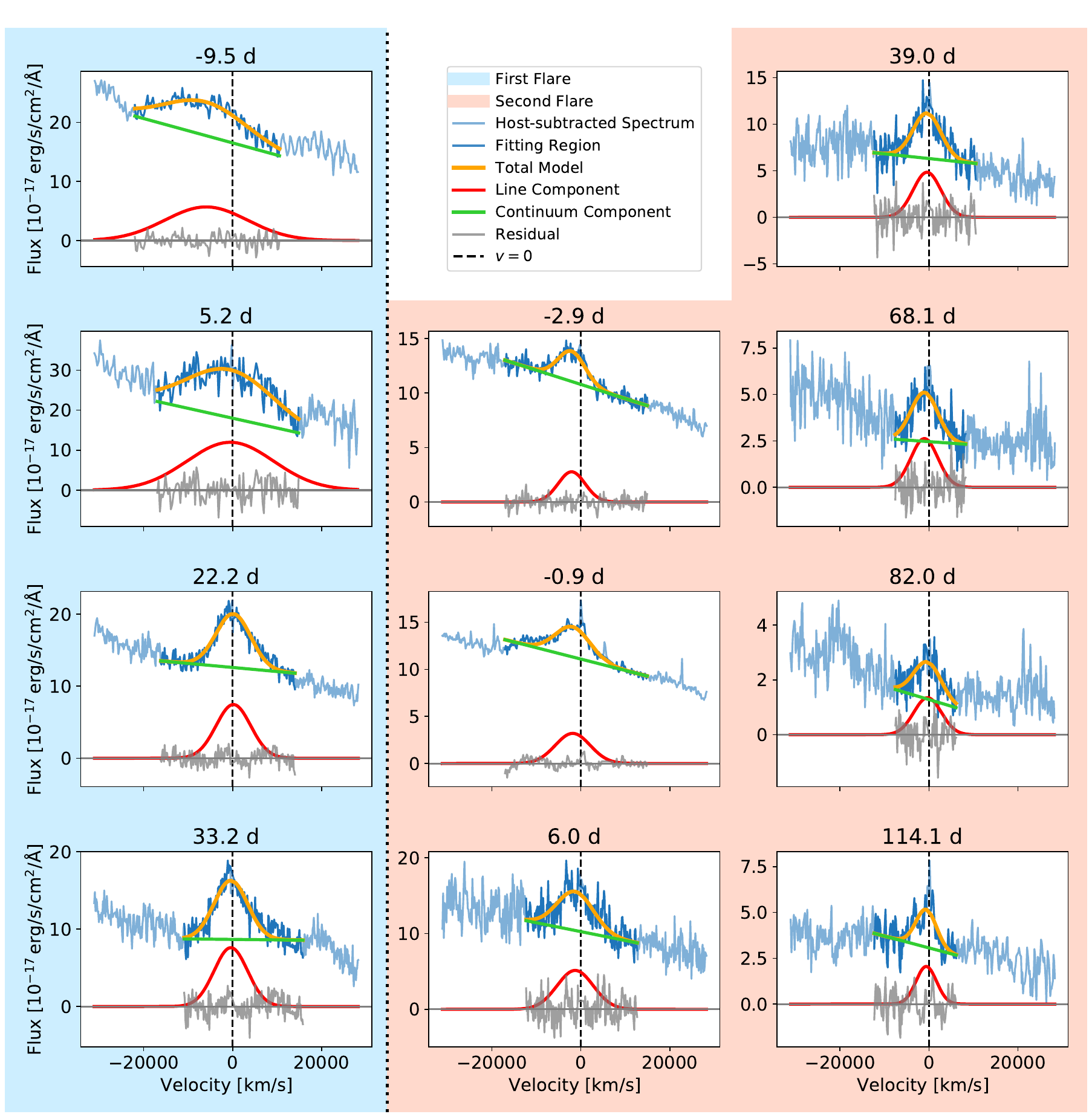}
  \caption{H$\alpha$ line-profile fits at all epochs. Each panel shows the host-subtracted spectrum (blue), the best-fit total model (orange), and its decomposition into the continuum (green) and emission-line (red) components. The fit residuals are shown in gray, and zero velocity is marked by the vertical black line. Epochs from the first flare are shown on the left in blue, while those from the second flare are shown on the right in red.}
  \label{fig:Halphaproof}
\end{figure*}

To measure the line profiles of H$\alpha$, we modeled the feature using a Gaussian profile with a free centroid and a linear continuum. For each spectrum, we fit this model in velocity space to extract the FWHM, velocity offset, and line luminosity. The parameter estimation was performed using the nested sampling Monte Carlo algorithm implemented in the \verb+UltraNest+ package, which yields the best-fit parameters and their 1-$\sigma$ uncertainties. The resulting FWHM, velocity offset, and luminosity measurements for all 11 spectra are listed in Table~\ref{Tab:Halpha}. The fits and residuals for each epoch are shown in Figure~\ref{fig:Halphaproof}.

These values are compared with a sample of spectroscopically characterized TDEs from \citet{charalampopoulos2022detailed}. The results are visualized in Figure \ref{fig:halpha}. Because H$\alpha$ luminosity strongly correlates with bolometric luminosity, we normalize each event by its peak bolometric luminosity to enable meaningful comparison.

\startlongtable
\begin{deluxetable*}{cccccccc}
\tablecaption{H$\alpha$ Emission Feature}
\tablehead{
\colhead{MJD} &
\colhead{Phase} &
\colhead{FWHM} &
\colhead{Uncertainty} &
\colhead{Offset} &
\colhead{Uncertainty} &
\colhead{Luminosity} &
\colhead{Uncertainty} \\
\colhead{(d)} &
\colhead{(d)} &
\colhead{(km\,s$^{-1}$)} &
\colhead{(km\,s$^{-1}$)} &
\colhead{(km\,s$^{-1}$)} &
\colhead{(km\,s$^{-1}$)} &
\colhead{($10^{40}$ erg\,s$^{-1}$)} &
\colhead{($10^{40}$ erg\,s$^{-1}$)}
}
\startdata
59631.6 & $-$9.5  & 22000 & 3900 & $-$6000 & 1300 & 5.40  & 1.34 \\
59646.4 & 5.2     & 22900 & 2900 & $-$600  & 1500 & 11.78 & 2.15 \\
59663.3 & 22.2    & 9160  & 350  & 140     & 110  & 2.93  & 0.14 \\
59674.4 & 33.2    & 8860  & 850  & $-$430  & 180  & 2.89  & 0.31 \\
\hline
60345.5 & $-$2.9  & 7160  & 300  & $-$2080 & 120  & 0.85  & 0.05 \\
60347.4 & $-$0.9  & 9210  & 460  & $-$1830 & 130  & 1.26  & 0.07 \\
60354.4 & 6.0     & 9600  & 2400 & $-$1260 & 580  & 2.10  & 0.61 \\
60387.4 & 39.0    & 7380  & 930  & $-$440  & 270  & 1.53  & 0.23 \\
60416.4 & 68.1    & 7200  & 1900 & $-$1050 & 450  & 0.81  & 0.27 \\
60430.4 & 82.0    & 7100  & 3700 & $-$330  & 760  & 0.41  & 0.24 \\
60462.4 & 114.1   & 5500  & 1200 & $-$640  & 350  & 0.48  & 0.13 \\
\enddata
\label{Tab:Halpha}
\end{deluxetable*}

The emission features \ion{N}{3} at 4640~\AA\ and \ion{He}{2} at 4686~\AA\ are blended due to their intrinsic velocity width, making it difficult to separate them cleanly. To characterize this combined feature, we use two complementary fitting approaches: one to track total luminosity and the other to diagnose dominance. First, we fit the profile using two Gaussians with fixed centers at 4640~\AA\ and 4686~\AA\ and a shared width, and a linear continuum to model the blended emission. We then estimated the luminosity of the feature using a non-parametric direct integration method. The evolution of the luminosity ratio between the combined \ion{N}{3}/\ion{He}{2} feature and H$\alpha$ is shown in Figure~\ref{fig:NIII}. Second, we fit the same feature using a single Gaussian with a free centroid parameter to determine the line dominance. The centroid evolution is presented in Figure~\ref{fig:centroid}. We compared the luminosity ratio of \ion{N}{3} and \ion{He}{2} over H$\alpha$ with the TDE sample from the detailed spectroscopic study in \citet{charalampopoulos2022detailed}. 

\startlongtable
\begin{deluxetable*}{cccccc}
\tablecaption{NIII/HeII Emission Feature}
\tablehead{
\colhead{MJD (d)} & \colhead{Phase (d)} & Centroid (\AA) & \colhead{ Uncertainties (\AA)} & \colhead{$L$ ($10^{40}$ erg\,s$^{-1}$)} & \colhead{ Uncertainties ($10^{40}$ erg\,s$^{-1}$)}
}
\startdata
59631.6 & $-$9.5  & 4630.21 & 0.20  & 4.62 & 2.92 \\
59646.4 & 5.2     & 4632.81 & 2.64  & 6.30 & 2.97 \\
59663.3 & 22.2    & 4640.77 & 1.78  & 3.10 & 0.22 \\
59674.4 & 33.2    & 4639.45 & 4.12  & 1.91 & 0.29 \\
\hline
60345.5 & $-$2.9  & 4633.09 & 2.41  & 2.82 & 0.38 \\
60347.4 & $-$0.9  & 4637.52 & 2.55  & 3.25 & 0.24 \\
60354.4 & 6.0     & 4653.52 & 11.08 & 2.20 & 0.79 \\
60387.4 & 39.0    & 4670.14 & 8.61  & 1.37 & 0.52 \\
60416.4 & 68.1    & 4667.82 & 15.22 & 0.71 & 0.42 \\
60430.4 & 82.0    & 4676.13 & 11.18 & 1.05 & 0.52 \\
60462.4 & 114.1   & 4666.64 & 19.05 & 0.38 & 0.38 \\
\enddata
\label{Tab:NIIIHeII}
\end{deluxetable*}

\subsection{H$\alpha$ Behavior}
\begin{figure}[h!]
  \centering
  \includegraphics[width=\columnwidth]{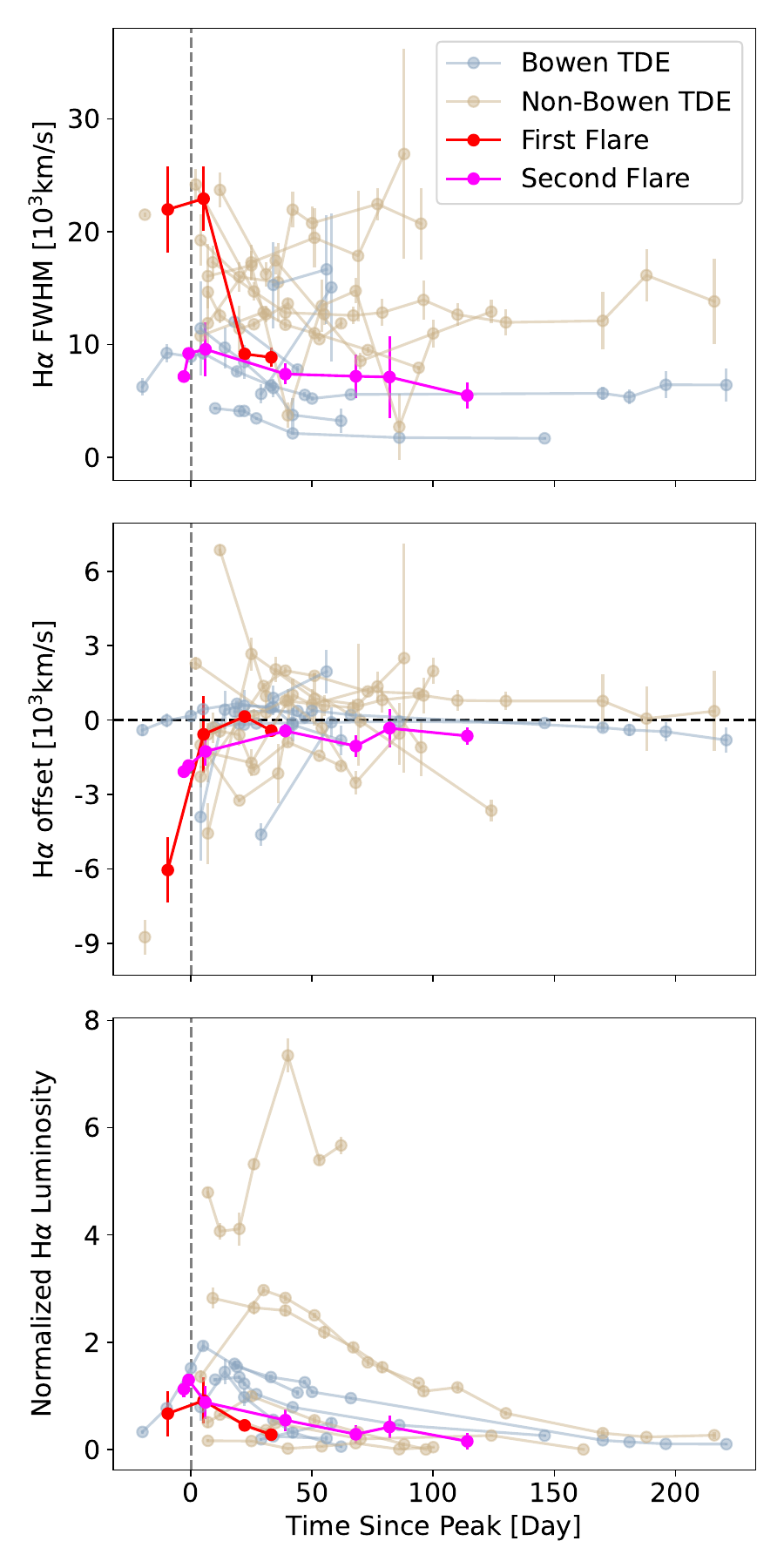}
  \caption{Evolution of the H$\alpha$ FWHM, velocity offset, and normalized luminosity in TDE\,2022dbl (top to bottom panels), compared to those of TDEs with and without Bowen \ion{N}{3} features. Red and magenta points show the first and second flares of TDE\,2022dbl, compared with light blue (Bowen) and gold (non-Bowen) TDEs from \citet{charalampopoulos2022detailed}. Luminosities are normalized by the peak bolometric luminosity of each event (scaled by $10^3$ for clarity). The gray dashed vertical line marks the peak time.}
  \label{fig:halpha}
\end{figure}

As shown in the top panel of Figure~\ref{fig:halpha}, the H$\alpha$ FWHM in both flares rises to a maximum a few days after the optical peak and subsequently declines. Despite this common evolution, the near-peak line widths differ significantly between the two flares. The first flare displays a broader FWHM of 22,000~$\pm$~3,900 and 22,900~$\pm$~2,900~km~s$^{-1}$ at $-9.5$ and $+5.2$ days relative to the first peak, respectively. In contrast, the second flare reaches only $9,210 \pm 460$ and $9,600 \pm 2,400$~km~s$^{-1}$ at $-0.9$ and $+6.0$ days from the second peak, approximately half the near-peak values of the first flare.

Excluding the exceptionally broad near-first-peak spectra, both flares are broadly consistent with the FWHM distribution of the Bowen TDE sample. The large FWHM measured during the first flare may be partially attributed to optical depth effects. This result contrasts with the findings of \citet{makrygianni2025double}, who reported a peak H$\alpha$ FWHM of only $\sim$5,000~km~s$^{-1}$ for TDE\,2022dbl, based on a different spectroscopic dataset.

The velocity offsets of both flares show significant blueshifts at early times that gradually diminish. All pre-peak spectra exhibit blueshifts of several thousand~km~s$^{-1}$, with the first flare showing approximately three times the offset of the second. In particular, the first flare has a velocity offset of $-6,000~\pm~1,300$~km~s$^{-1}$ at $-9.5$ days relative to peak, while the second flare shows offsets of $-2,080~\pm~120$ and $-1,830~\pm~130$~km~s$^{-1}$ at $-2.9$ and $-0.9$ days, respectively. The subsequent decline in blueshift is consistent with a scenario in which an early-time, optically thick outflow becomes progressively transparent.

The bottom panel of Figure~\ref{fig:halpha} shows the evolution of the normalized H$\alpha$ luminosity. Both flares of TDE\,2022dbl can be characterized by a near-peak rise followed by a steady decline. Their normalized H$\alpha$ luminosities and overall evolution are consistent with those observed in Bowen TDEs, whose H$\alpha$ luminosities generally lie toward the lower end of the full TDE sample.

For comparison, the second flare of TDE\,2020vdq exhibited ultrabroad Balmer profiles with ${\rm FWHM}\sim0.084\,c$ that decreased slightly with time, while the line centroids became increasingly blueshifted \citep{somalwar2023first}. The peak H$\alpha$ width in the first flare of TDE\,2022dbl is comparable ($\sim0.07\,c$), whereas the second flare is narrower by roughly a factor of two. Both events show a decline in FWHM with time. In contrast to TDE\,2020vdq, however, TDE\,2022dbl displays a decreasing blueshift as the transient evolves.

Although the Gaussian model does not fully reproduce the H$\alpha$ line profiles, it provides a useful characterization of their overall evolution. In several epochs (e.g., +5.2 and +22.2 days for the first flare, +39.0 and +68.1 days for the second flare), the H$\alpha$ emission exhibits an enhanced red wing and a suppressed blue wing. The outflow kinematics, optical depth, and electron scattering can produce a blueshifted peak, whereas photons doing adiabatic work on the expanding gas lose energy, decrease in frequency, and therefore stretch the red wing, making the emission line profile asymmetric \citep{roth2018sets}.

\subsection{\ion{N}{3}/\ion{He}{2} Behavior}
\begin{figure}[h]
  \centering
  \includegraphics[width=\columnwidth]{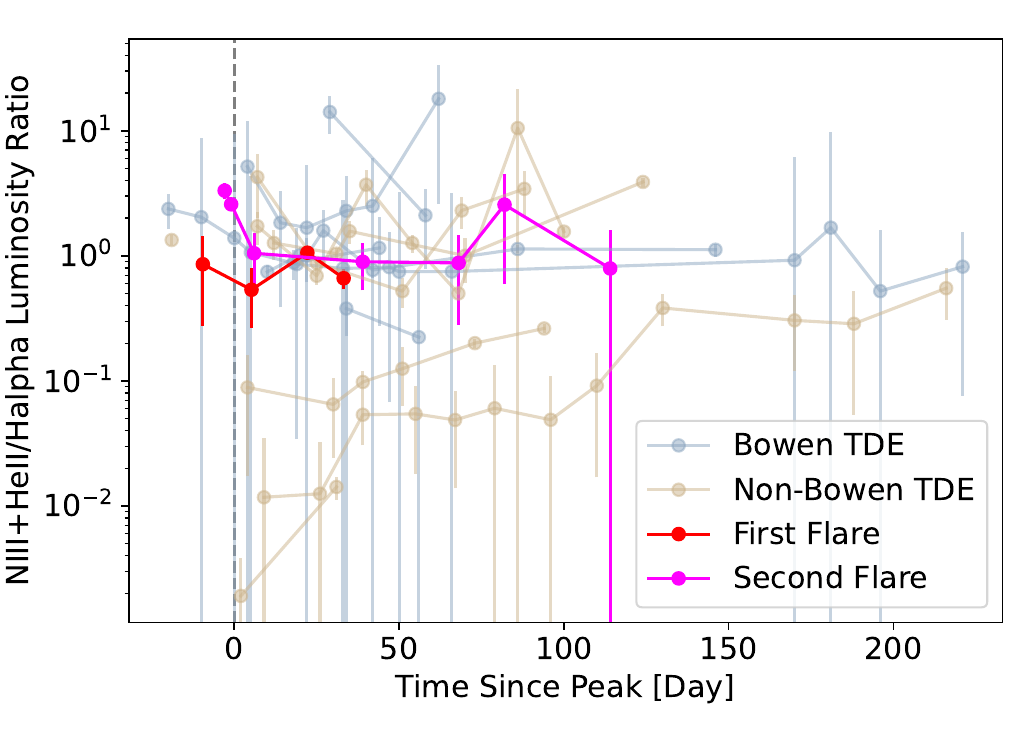}
  \caption{Evolution of the luminosity ratio between the blended \ion{N}{3}+\ion{He}{2} emission and H$\alpha$ for the first flare (red) and second flare (magenta) of TDE\,2022dbl, compared with Bowen (light blue) and non-Bowen TDEs (gold). The gray dashed vertical line marks the peak time.}
  \label{fig:NIII}
\end{figure}

Figure~\ref{fig:NIII} shows the evolution of the blended 4640/4686~\AA\ emission relative to H$\alpha$ across both flares of TDE\,2022dbl. For the non-Bowen TDE comparison sample, which is selected to have detections of both \ion{He}{2} and H$\alpha$, the plotted ratio is the \ion{He}{2} luminosity divided by the H$\alpha$ luminosity. For Bowen TDEs, the ratio reflects the combined luminosity of \ion{N}{3} and \ion{He}{2} relative to H$\alpha$. Within $\sim50$ days of peak, TDE\,2022dbl follows the general trend observed among Bowen TDEs. At later times, the second flare remains spectroscopically detected and maintains a relatively high ratio compared with most Bowen TDEs at similar phases, including a possible enhancement around $\sim80$ days after peak. However, this comparison is limited by the small number of events with late-time spectral coverage in the reference sample.

\begin{figure}[h]
  \centering
  \includegraphics[width=\columnwidth]{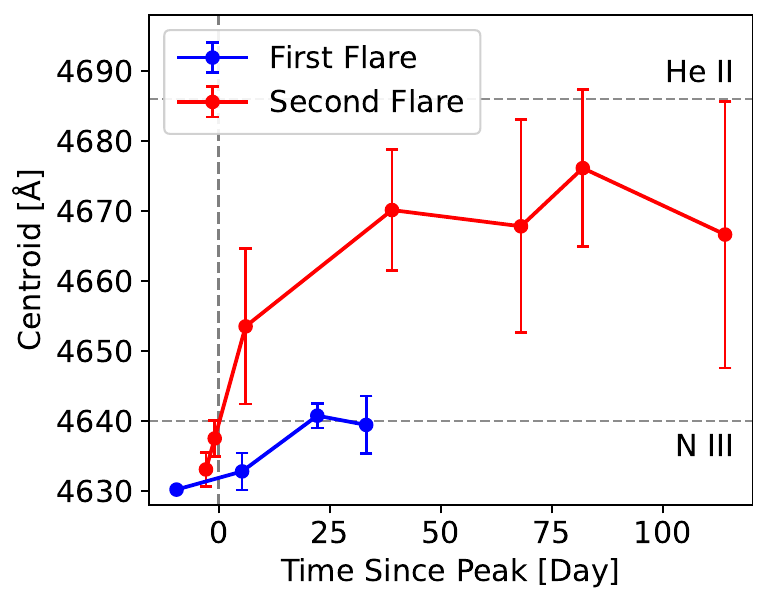}
  \caption{Evolution of the centroid of the blended \ion{N}{3} and \ion{He}{2} feature in TDE\,2022dbl. The dashed horizontal lines mark the rest-frame \ion{He}{2} and \ion{N}{3} wavelengths. The dashed vertical line marks the peak time.}
  \label{fig:centroid}
\end{figure}

As shown in Figure~\ref{fig:centroid}, the centroid evolution of the blended 4640/4686~\AA\ feature reveals distinct behaviors across the two flares. During the first flare, the centroids remain close to 4640~\AA, indicating that the emission is dominated by \ion{N}{3}. In contrast, the second flare shows a gradual shift in centroid from 4640~\AA\ toward 4686~\AA, suggesting an initial dominance by \ion{N}{3} followed by increasing contribution from \ion{He}{2}. This interpretation is supported by the detection of the 4100~\AA\ \ion{N}{3}-only feature in early-time spectra of the second flare. By $+68.1$~days and later, the \ion{N}{3} features have faded significantly. 

\begin{figure}[h]
  \centering
  \includegraphics[width=\columnwidth]{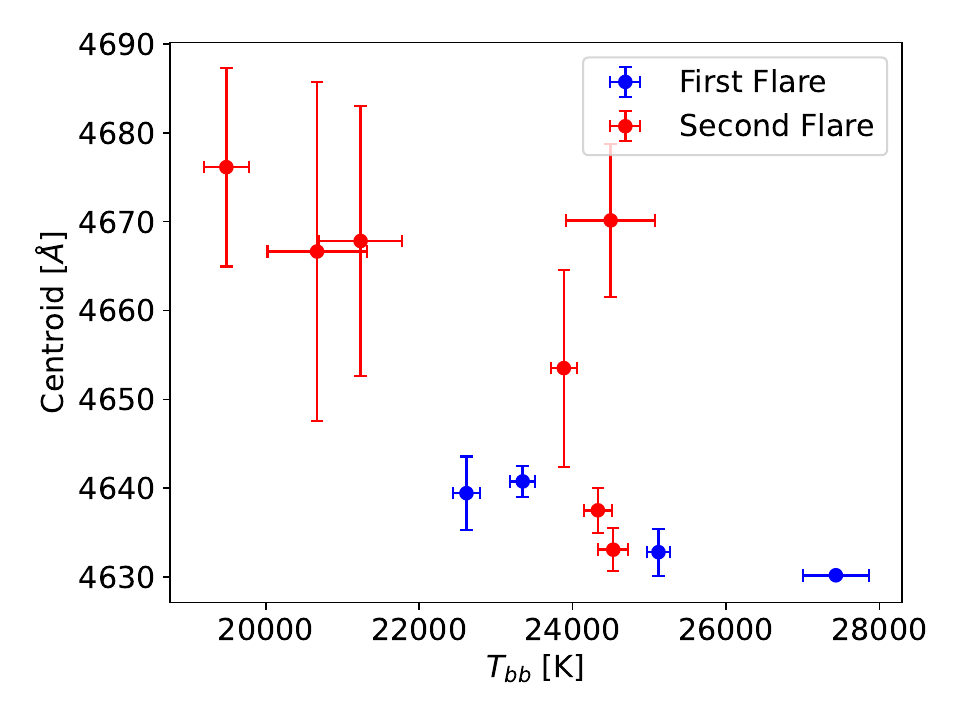}
  \caption{Scatter plot between centroid of \ion{N}{3} and \ion{He}{2} combined feature and the blackbody temperature in TDE\,2022dbl.}
  \label{fig:centroidTT}
\end{figure}

A plausible explanation is a decreasing level of ionization or efficiency of the Bowen mechanism as the blackbody temperature and UV flux drop. Indeed, Figure~\ref{fig:centroidTT} suggests a negative correlation between centroid position and $T_{\rm bb}$ during the second flare. However, the lack of centroid evolution in the first flare despite a similar temperature decline may indicate that ionization alone does not fully regulate line dominance. 
 
TDE\,2018dyb and ASASSN-14li are initially \ion{N}{3}-dominated, with the line ratio \ion{He}{2}/\ion{N}{3} increasing with time, while iPTF15af is \ion{N}{3}-dominated at $\sim25$ days \citep{leloudas2019spectral}. The evolution of TDE\,2018dyb and ASASSN-14li is consistent with that observed during the second flare of TDE\,2022dbl, while the first flare follows the behavior of iPTF15af more closely. \citet{leloudas2019spectral} also attributed the increasing relative contribution of \ion{He}{2} emission to changes in the ionization state, consistent with our interpretation.

The luminosity ratio of the blended \ion{N}{3}+\ion{He}{2} emission to H$\alpha$ emission first decreases and then increases, while the centroid of the blend shifts from \ion{N}{3} to \ion{He}{2} dominance. Together, these trends indicate that the \ion{He}{2} luminosity grows at late times. The opposite trend of \ion{He}{2} relative to H$\alpha$ suggests that the two lines originate from physically distinct regions: \ion{He}{2} likely forms in a denser, more highly ionized inner layer, while H$\alpha$ originates in a more extended, less ionized outer region.

\section{Discussion} 
\label{sec:disc}
Prior work discussed various possibilities for this event in detail, including gravitational lensing, the disruption of two different stars, and the capture of the same star using the Hills mechanism, concluding that the most likely scenario was that the same star was disrupted twice \citep{lin2024unluckiest, hinkle2024double, makrygianni2025double, hinkle2026disappearingactconstraintsmissing}. We adopt this framework in the following discussion.

The two flares of TDE\,2022dbl share many properties both in terms of continuum and line features. Throughout their evolution, the corresponding values of $T_{\rm bb}$, $R_{\rm bb}$, and $L_{\rm bb}$ remain within approximately a factor of three of each other. The spectra of both flares exhibit similar line components, and their features are broadly consistent with the Bowen-TDE sample, especially at late times. The two flares of TDE\,2022dbl have peak luminosities of $\log(L_{\rm bb}/{\rm erg~s^{-1}})=43.80$ and $43.32$, placing both at the lower end of the TDE luminosity distribution, which spans a wide range of $\simeq43.3$--$45.8$ \citep{yao2023tidal}. In addition, the comparable thermal properties and spectral line behaviors are consistent with the same star being disrupted by the same SMBH.

However, the two flares differ in their detailed spectral and light-curve properties, especially near the peak. The bolometric light curve behavior in Figure \ref{fig:bolometric} shows that the first flare has a relatively sharp peak, while the second flare has a flatter peak at a lower luminosity. Figure \ref{fig:riseanddecline} shows that the second flare has a longer decline time. Moreover, the spectral lines during the first flare have larger FWHM and blueshifts than during the second flare near peak. These differences may suggest that the second flare undergoes a smoother, more prolonged release of energy rather than directly tracing the fallback rate.

We argue that a pre-existing accretion disk may exist at the time of the second flare. The \swift-uvw2 detection at $\sim595$ days after the first peak (S/N$=3.04$) suggests persistent late-time UV excess, while the increase in X-ray brightness between the two flares provides further evidence for ongoing accretion \citep{hinkle2024double}. If debris from the second stellar stripping interacts with this pre-existing disk, it may dissipate energy through stream--disk interactions. The observed emission could therefore evolve on viscous and diffusion timescales rather than directly following the fallback rate. Disk-regulated emission would produce a smoother release of energy, naturally explaining the broader peak and longer decline timescale of the second flare. Moreover, the lower FWHM of the spectral lines near the second peak could reflect a less violent outflow, whereas the lower peak luminosity may result in reduced radiation pressure and a lower optical depth, contributing to the smaller near-peak blueshift observed in the second flare. Additionally, the thermal, photometric, and spectroscopic differences between the two flares become smaller with time, possibly indicating convergence toward a common accretion configuration at late times.

We note that a shallower encounter could also increase the fallback timescale and reduce the peak luminosity. This potential scenario can also explain the lower FWHM and blueshift in the second flare, as less stripped mass produces a weaker outflow velocity and lower optical depth. However, fallback scaling alone does not naturally account for changes in the morphology near the peak of the light curve. A shallower encounter would also increase the likelihood that the star survives to undergo another disruption, although the recurrence interval would remain uncertain and depend on the magnitude and asymmetry of the kick.

Both flares of TDE\,2022dbl are broadly consistent with the observed optical–UV TDE population. Without knowledge of the recurrence, both flares would likely be classified as ordinary TDEs. Thus, observations of a single flare may not uniquely determine whether the underlying disruption was full or partial. Some isolated TDEs could therefore be pTDEs whose next flare has not been observed yet \citep{makrygianni2025double}. Consequently, the confirmation of pTDEs still mainly depends on observation of multiple flares \citep{hinkle2024double, makrygianni2025double}.

\section{Where is the Third Flare?}\label{sec:thirdflare}
If TDE\,2022dbl maintained its $\sim700$-day recurrence period, a third flare was expected near February 2026 \citep{hinkle2024double}. However, no comparable flare has been observed as of mid-2026. In this section, we consider possible explanations for the missing third flare.

One possible explanation is that the two flares were produced by two independent TDEs. However, even accounting for the enhanced TDE rate in post-starburst galaxies, the short recurrence timescale, and the luminosity distribution, this scenario has a probability below 0.5\% \citep{makrygianni2025double, hinkle2026disappearingactconstraintsmissing}. Avoiding this low probability would require twin-flare hosts to have unusually high TDE rates drawn from an extremely broad, strongly right-skewed rate distribution, inconsistent with observations \citep{hinkle2026disappearingactconstraintsmissing}. A double TDE involving a stellar binary is another possibility, but \citet{hinkle2026disappearingactconstraintsmissing} find that sequential disruptions occur too quickly ($<150$ days), whereas the return of a captured companion occurs too slowly (median 50 yrs). We thus consider alternatives in which both flares arise from repeated encounters of a single star with the SMBH.

One possibility is that the system naturally produces only two observable flares. \citet{bandopadhyay2024repeating} simulated the repeated disruption of a $1\,M_{\odot}$ zero-age main-sequence star by a $10^6\,M_{\odot}$ SMBH. In their model, a deep initial encounter with $\beta=1.5$ leaves behind a stripped remnant that is completely disrupted on the next pericenter passage. Because the second disruption has a lower peak fallback rate than the first, this scenario naturally produces a fainter second flare and no subsequent flare, consistent with the behavior of TDE\,2022dbl if the luminosity traces the fallback or accretion rate \citep{bandopadhyay2024repeating}. We note, however, that TDE\,2020vdq, another ``two flares and done'' candidate \citep{hinkle2026disappearingactconstraintsmissing}, showed the opposite evolution, with a brighter second flare than the first \citep{somalwar2023first}. Building on these hydrodynamical results, \citet{hinkle2026disappearingactconstraintsmissing} used a semi-analytic model to explore this ``two flares and done'' outcome over a broader range of stellar structures and orbital parameters. They found that such an outcome is plausible for a main-sequence star placed on a bound orbit after a deep initial encounter close to the full-disruption threshold.

Here, we consider an additional interpretation: the remnant may survive the second disruption but be placed on a longer-period bound orbit, delaying the next flare beyond the expected window. One potential explanation is that during the disruption, additional energy is deposited into the orbit of the surviving stellar core, lengthening its return period. One commonly discussed mechanism is recoil from asymmetric mass loss \citep{manukian2013turbovelocity, gafton2015relativistic, chen2024fate}. More recently, \citet{coughlin2025s} argued that the primary mechanism responsible for the positive-energy kick is ballistic core reformation, in which the reformed core inherits positive orbital energy from the sheared debris stream.

For a $M_{\rm BH}\sim10^6\,M_{\odot}$, the tidal radius is $r_t\simeq47\,r_g$, so relativistic effects should not change the result by orders of magnitude. We therefore assume Keplerian orbits and use non-relativistic calculations in this section. We adopt the black hole mass derived in Section~\ref{sec:SMBHmass}, $M_{\rm BH}=10^{6.01}M_{\odot}$, and assume a solar-mass star with $M_{\star}=1\,M_{\odot}$. For an orbital period of $P_0\simeq700$ days between the first and second disruptions, Kepler's third law gives
\begin{equation}
a =
\left(
\frac{G M_{\mathrm{BH}} P_0^2}{4\pi^2}
\right)^{1/3}
= 2 \times 10^{15} \ \mathrm{cm}.
\end{equation}
The Keplerian specific orbital energy for the star before the second disruption is then
\begin{equation}
E =
-\frac{G M_{\mathrm{BH}}}{2a}
=
-3.0 \times 10^{16} \ \mathrm{cm^2 \ s^{-2}}.
\end{equation}
Following the convention of \citet{coughlin2025s}, we define $v_{\rm kick}$ as the energy-equivalent kick speed as $v_{\rm kick}\equiv\sqrt{2\Delta E}$, where $\Delta E$ is the change in orbital energy as
\begin{equation}
\Delta E
=
3.0 \times 10^{16}
\left[
1 -
\left(
\frac{P_0}{P_1}
\right)^{2/3}
\right] \mathrm{cm^2 \ s^{-2}},
\end{equation}
where $P_0\simeq700$ days is the period between the first and second disruptions, and $P_1$ is the period between the second flare and a possible third flare.

However, the kick speed cannot grow without bound, so this mechanism cannot delay the next return arbitrarily far. In the ballistic core-reformation model, the maximum kick is set by the saturation of the energy inherited by the reformed core. Once the outward, higher-energy portion of the debris approaches homologous expansion, the density profile no longer shifts its peak toward higher energies. \citet{coughlin2025s} estimate this limiting kick speed as
\begin{equation}
v_{\mathrm{kick,max}}
\simeq
0.4\,v_{\mathrm{esc}}
\left(
\frac{M_{\mathrm{BH}}}{M_{\star}}
\right)^{1/6},
\end{equation}
where $v_{\rm esc}\equiv\sqrt{2GM_\star/R_\star}$ is the stellar escape velocity. For our adopted parameters, this gives $v_{\rm kick,max}\sim2480$ km s$^{-1}$. 

To unbind the remnant orbit from the SMBH, the required kick speed is $v_{\rm kick} \sim 2400$ km s$^{-1}$. Complete ejection is therefore possible only near the maximum attainable kick speed, and we consider this scenario unlikely.

The core reformation time $t_{\rm reform}$ can be inferred within the ballistic core-reformation model of
\begin{equation}
t_{\rm reform}
=
\beta^{-3/2}
\sqrt{\frac{R_\star^3}{G M_\star}}
\left[
\frac{v_{\rm kick}/v_{\rm esc}}
{0.533\sqrt{\kappa}}
\right]^3,
\end{equation}
where we adopt $\beta=0.9$, appropriate for an encounter near the disruption threshold, and $\kappa\simeq0.095$ for a $5/3$ polytrope model of the star, with $\kappa$ parameterizing the dependence on stellar structure and encounter geometry  \citep{coughlin2025s}. Assuming a solar-like star, this becomes 
\begin{equation}
    t_{\rm reform} \simeq 4.87
\left(
\frac{v_{\rm kick}}{v_{\rm esc}}
\right)^3
\,{\rm days}.
\end{equation}

The ballistic core-reformation scenario can naturally produce a modest orbital delay. Lengthening the recurrence period from $P_0\simeq700$ days to $P_1\simeq970$ days (as of October 2026) requires an energy-equivalent kick of $v_{\rm kick}\sim1050$ km s$^{-1}$, corresponding to $v_{\rm kick}/v_{\rm esc}\sim1.7$ and $t_{\rm reform} \sim 24$ days. $t_{\rm reform}$ is difficult to predict from first principles, as it is highly sensitive to the self-gravity evolution and perturbations within the stellar debris \citep{coughlin2023, coughlin2025s}. Although $t_{\rm reform}$ as long as 24 days has not yet been directly demonstrated in hydrodynamic simulations, the required kick remains well below the maximum allowed by the ballistic core-reformation model. Thus, the missing flare at the strictly periodic return time is still consistent with a surviving remnant on a longer-period bound orbit.

To illustrate a range of possible delays, we consider return periods of 3 and 5 years. A 3-year recurrence period requires $v_{\rm kick}\sim1210$ km s$^{-1}$ and $t_{\rm reform}\sim37$ days, which remains plausible within the model. A more extreme 5-year period requires $v_{\rm kick}\sim1650$ km s$^{-1}$ and $t_{\rm reform}\sim93$ days. This case lies in a regime where the model has not been well tested. Moreover, for an initially solar-mass star, the reformed surviving core in this case would contain only $\sim5\,M_{\rm Jup}$ of material, where $M_{\rm Jup}$ is the mass of Jupiter. Even if this remnant returns, the resulting flare could therefore be substantially fainter than the first two and may fall below current detection limits. Continued monitoring of TDE\,2022dbl is therefore important, both for gaining invaluable pTDE observational data and for constraining theoretical models of rare repeating pTDEs.

We note that the core-reformation mechanism is not expected to operate equally in all pTDE systems. Some rpTDEs, such as ASASSN-14ko and AT\,2023uqm, show multiple optical flares with relatively stable recurrence times. The relevance of this mechanism may therefore depend on configurations of the systems. Under the pTDE interpretation, hydrodynamic modeling of ASASSN-14ko favors a moderately massive evolved star, whose internal structure consists of a dense core surrounded by a loosely bound envelope \citep{Liu2023TDE, bandopadhyay2024repeating}. Such a configuration can survive repeated grazing encounters with the SMBH while losing only a small amount of envelope material in each passage. In this case, the stellar core may remain largely intact, so ballistic core reformation and the orbital-energy kick may be less important, making a measurable orbital delay less likely. Similarly, AT\,2023uqm's double-peaked light curves have been interpreted as evidence for a giant-star progenitor \citep{wang2025star, Liu2025rpTDE}, in which case the same argument would apply. However, its observed runaway mass loss may suggest a main-sequence progenitor \citep{wang2025star}, for which core reformation could become important. If future flares of AT\,2023uqm deviate from the expected recurrence time, this would provide a useful observational test of the ballistic core-reformation model.

\section{Conclusions} \label{sec:conc}
In this work, we have reexamined TDE\,2022dbl using improved photometric data, an expanded spectroscopic data set, and a new high-resolution host-galaxy spectrum. We reanalyze the \swift-UVOT data incorporating the instrument's wide PSF and trailing problem. We also present a new optical spectroscopic sequence spanning both flares, enabling a systematic comparison of their spectral evolution. In addition, we report a new measurement of the host-galaxy stellar velocity dispersion obtained with high-resolution KeckII/ESI. With these improved measurements and expanded observations, we identify both the similarities and differences between the two flares and use them to investigate the physical evolution of the system and the nature of the missing third flare.

From the ESI spectrum of the host galaxy, we measure a stellar velocity dispersion of $\sigma = 53.3\pm2.1\ {\rm km,s^{-1}}$. Applying the $M-\sigma$ relation from \citet{greene2020intermediate}, we infer a central black hole mass of $\log_{10}(M_{\rm BH}/M_\odot)=6.01\pm0.45$, which is consistent with the values adopted in previous work \citep{lin2024unluckiest, hinkle2024double, angus2026can, mummery2025optical}.

In agreement with \citet{lin2024unluckiest}, \citet{hinkle2024double}, and \citet{makrygianni2025double}, we find that the two flares are broadly similar. Their thermal properties remain within a factor of three. In particular, blackbody fits show that the temperatures of both flares decline from approximately $27{,}000$ K to $15{,}000$ K within $\sim100$ days after peak. The logarithmic peak luminosities (in erg~s$^{-1}$) of the first and second flares are 43.80 and 43.32, respectively, placing both events at the lower-luminosity end of the pTDE population. The two flares released comparable total energies, although the first flare, with an energy of $\sim2\times10^{50}$ erg, is approximately 35\% more energetic than the second. We detect late-time UV emission in the \swift-uvw2 band following both flares, consistent with the presence of an accretion disk that becomes more luminous after the second flare. We also analyzed 11 optical spectra spanning the two flares and identified prominent emission from H$\alpha$, \ion{N}{3}, and \ion{He}{2}, firmly placing TDE\,2022dbl within the Bowen TDE class. H$\alpha$ and blended Bowen features follow a common temporal pattern across the two flares, particularly away from peak. Together, the comparable photometric and spectroscopic evolution between the two flares suggests that TDE\,2022dbl is a very promising partial TDE candidate.

Despite these broad similarities, the two flares also exhibit several notable differences. The first flare shows a sharp peak, while the second flare displays a plateau-like feature near the peak. By fitting both light curves with a power-law rise and exponential-decline model, we find that the second flare has a longer decline timescale, potentially indicating a reduced fallback rate or more strongly regulated emission. Although the two flares exhibit broadly consistent spectral-line evolution, the peak FWHM of H$\alpha$ during the first flare is approximately two times larger than that measured during the second. At early times, the emission feature near 4650~\AA\ is dominated by \ion{N}{3} $\lambda$4640 over \ion{He}{2} $\lambda$4686 in both flares, but in the second flare, the dominance shifts to \ion{He}{2} at about 68 days after the peak, suggesting evolving ionization conditions within the line-forming region. The \ion{He}{2} emission may arise from a denser and more highly ionized inner region than the H$\alpha$ emission. We suggest that a long-lived accretion disk formed during the first flare and interacted with debris from the second encounter. This may have regulated the energy release, naturally producing the plateau-like peak, longer light-curve timescales, and less extreme spectroscopic velocities observed during the second flare. Alternative explanations remain possible.

As of October 2026, a third flare has not occurred at the predicted time. Previous work has considered a ``two-flares-and-done scenario'' produced by a main sequence star after a deep initial encounter \citep{bandopadhyay2024repeating,hinkle2026disappearingactconstraintsmissing}. In this work, we consider an additional possible interpretation of the missing third flare, that ballistic core reformation imparts a positive-energy kick to the surviving stellar remnant, delaying its return to the pericenter \citep{coughlin2025s}. Delaying the recurrence to the current non-detection time requires $v_{\rm kick}\sim865$ km s$^{-1}$, corresponding to a core-reformation time of $\sim11$ days, which lies within the range expected by this model. In contrast, fully unbinding the remnant requires $v_{\rm kick}\sim2400$ km s$^{-1}$, or $v_{\rm kick}/v_{\rm esc}\sim3.9$, which is close to our expected maximum kick speed. Such a large kick would likely require a deep partial disruption close to the complete-disruption threshold. Thus, while ejection remains possible, a delayed bound return is more likely. Within this framework, a third flare occurring approximately three years after the second remains plausible, whereas a delay of five years would be substantially less likely to be observed and would extend into a regime not yet tested by the model.

Continued photometric and spectroscopic monitoring of TDE\,2022dbl and similar systems will be essential for characterizing the long-term behavior of pTDEs. This may help distinguish partial from full TDEs and study the physical mechanisms behind them. In addition, whether and when the third flare occurs would directly test theoretical models, providing key constraints on the nature of this rare and intriguing system.

\vspace{5mm}
\facilities{\swift-UVOT, \swift-XRT, ZTF, ATLAS, Shane (Kast Double spectrograph), Keck:I (LRIS), Keck:II (ESI), and LDT (DeVeny)}


\software{Matplotlib \citep{Hunter2007matplotlib}, NumPy \citep{harris2020array}, Astropy \citep{astropy:2013, astropy:2018, astropy:2022}, Prospector \citep{leja2017deriving, Johnson2021stellar}, pandas \citep{McKinney2010pd, pandas_software}, SciPy \citep{Virtanen2020scipy}, Fritz \citep{vanderWalt2019, Coughlin2023Fritz}}

\begin{acknowledgments}

X.G. acknowledges support from the 2024 Mathematics \& Physical Sciences (MPS) Scholars program at UC Berkeley and the Bensadoun 2025 summer undergraduate research funds. The authors gratefully acknowledge the generous philanthropic support of Cristina and Marc Bensadoun, who helped make this research possible. The authors thank V. Ravi for obtaining a Keck LRIS spectrum.

Based on observations obtained with the Samuel Oschin Telescope 48-inch and the 60-inch Telescope at the Palomar Observatory as part of the Zwicky Transient Facility project. ZTF is supported by the National Science Foundation under Grants No. AST-1440341, AST-2034437, and currently Award \#2407588. ZTF receives additional funding from the ZTF partnership. Current members include Caltech, USA; Caltech/IPAC, USA; University of Maryland, USA; University of California, Berkeley, USA; Cornell University, USA; Drexel University, USA; University of North Carolina at Chapel Hill, USA; Institute of Science and Technology, Austria; National Central University, Taiwan, German Center for Astrophysics, Germany, and OKC, University of Stockholm, Sweden. Operations are conducted by Caltech's Optical Observatory (COO), Caltech/IPAC, and the University of Washington at Seattle, USA. 

The Gordon and Betty Moore Foundation, through both the Data-Driven Investigator Program and a dedicated grant, provided critical funding for SkyPortal.
The ZTF forced-photometry service was funded under the Heising-Simons Foundation grant No. 12540303 (PI: Graham).

Some of the data presented herein were obtained at Keck Observatory, which is a private 501(c)3 non-profit organization operated as a scientific partnership among the California Institute of Technology, the University of California, and the National Aeronautics and Space Administration. The Observatory was made possible by the generous financial support of the W. M. Keck Foundation. 
The authors wish to recognize and acknowledge the very significant cultural role and reverence that the summit of Maunakea has always had within the Native Hawaiian community. We are most fortunate to have the opportunity to conduct observations from this mountain. 

A major upgrade of the Kast spectrograph on the Shane 3 m telescope at Lick Observatory, led by Brad Holden, was made possible through gifts from the Heising-Simons Foundation, William and Marina Kast, and the University of California Observatories. Research at Lick Observatory is partially supported by a generous gift from Google.

This work has made use of data from the Asteroid Terrestrial-impact Last Alert System (ATLAS) project. The ATLAS project is primarily funded to search for near earth asteroids through NASA grants NN12AR55G, 80NSSC18K0284, and 80NSSC18K1575; byproducts of the NEO search include images and catalogs from the survey area. This work was partially funded by Kepler/K2 grant J1944/80NSSC19K0112 and HST GO-15889, and STFC grants ST/T000198/1 and ST/S006109/1. The ATLAS science products have been made possible through the contributions of the University of Hawaii Institute for Astronomy, the Queen's University Belfast, the Space Telescope Science Institute, the South African Astronomical Observatory, and The Millennium Institute of Astrophysics (MAS), Chile.

This research has made use of the SVO Filter Profile Service "Carlos Rodrigo", funded by MCIN/AEI/10.13039/501100011033/ through grant PID2023-146210NB-I00. ChatGPT was used to suggest language edits for grammar, word choice, and sentence structure. The authors reviewed and selectively incorporated its suggestions.

\end{acknowledgments}

\appendix \label{sec:Appendix}
\restartappendixnumbering
\section{Supplementary Tables and Figures}
Here we provide supplementary tables and figures to support the host-galaxy, photometric, and spectroscopic analysis in the main text.

\begin{figure*}[t]
  \centering
  \includegraphics[width=\linewidth]{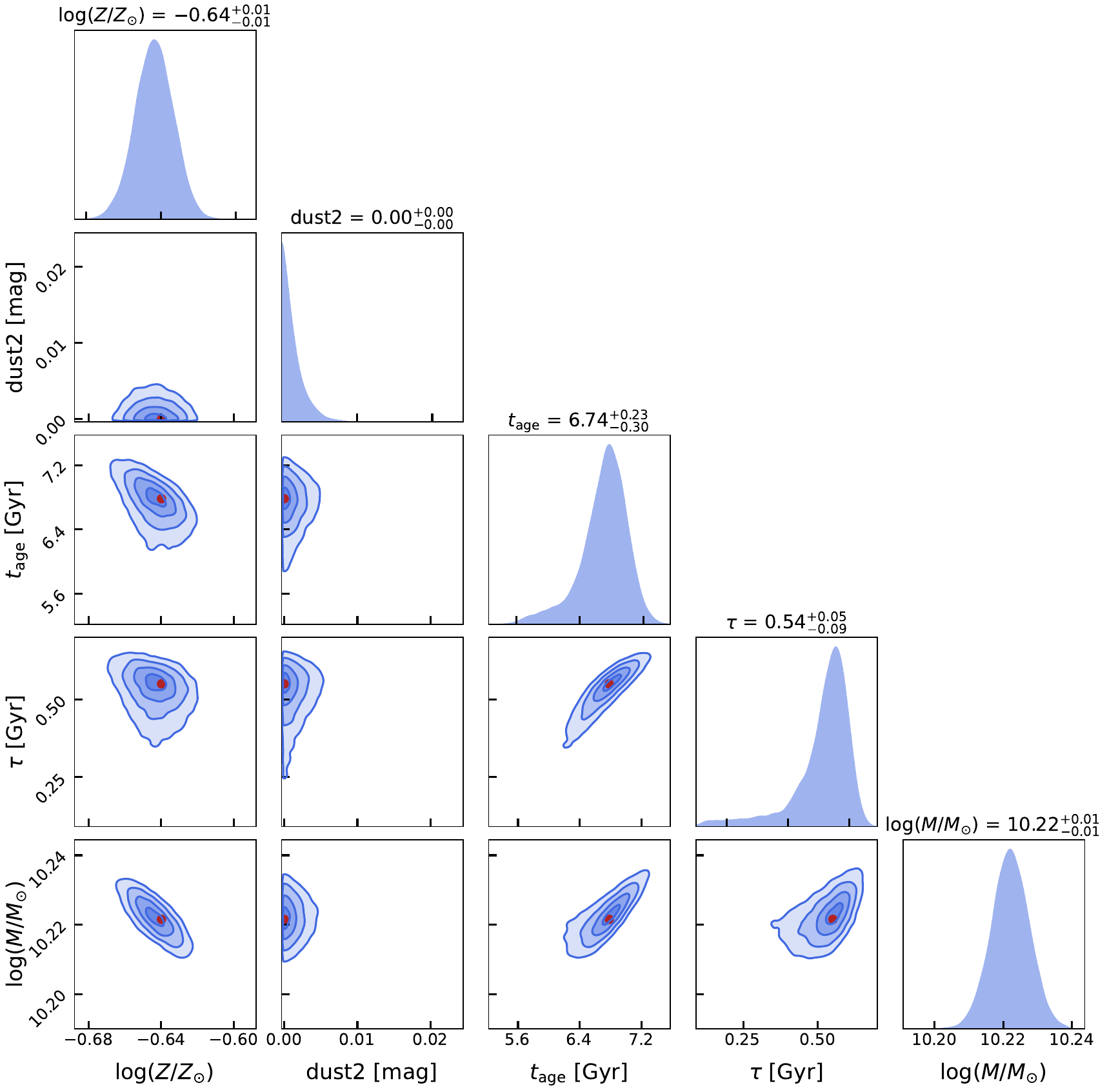}
  \caption{Corner plot of the posterior distributions from the \texttt{Prospector} host-galaxy fit. The diagonal panels exhibit the distribution of each parameter, whereas the contours in the off-diagonal panels represent the 0.5, 1, 1.5, and 2 $\sigma$ from innermost to outermost for the joint posterior distributions.}
  \label{fig:corner}
\end{figure*}

\begin{figure*}[htbp]
    \centering
    \begin{minipage}[b]{0.49\textwidth}
        \centering
        \includegraphics[width=\textwidth]{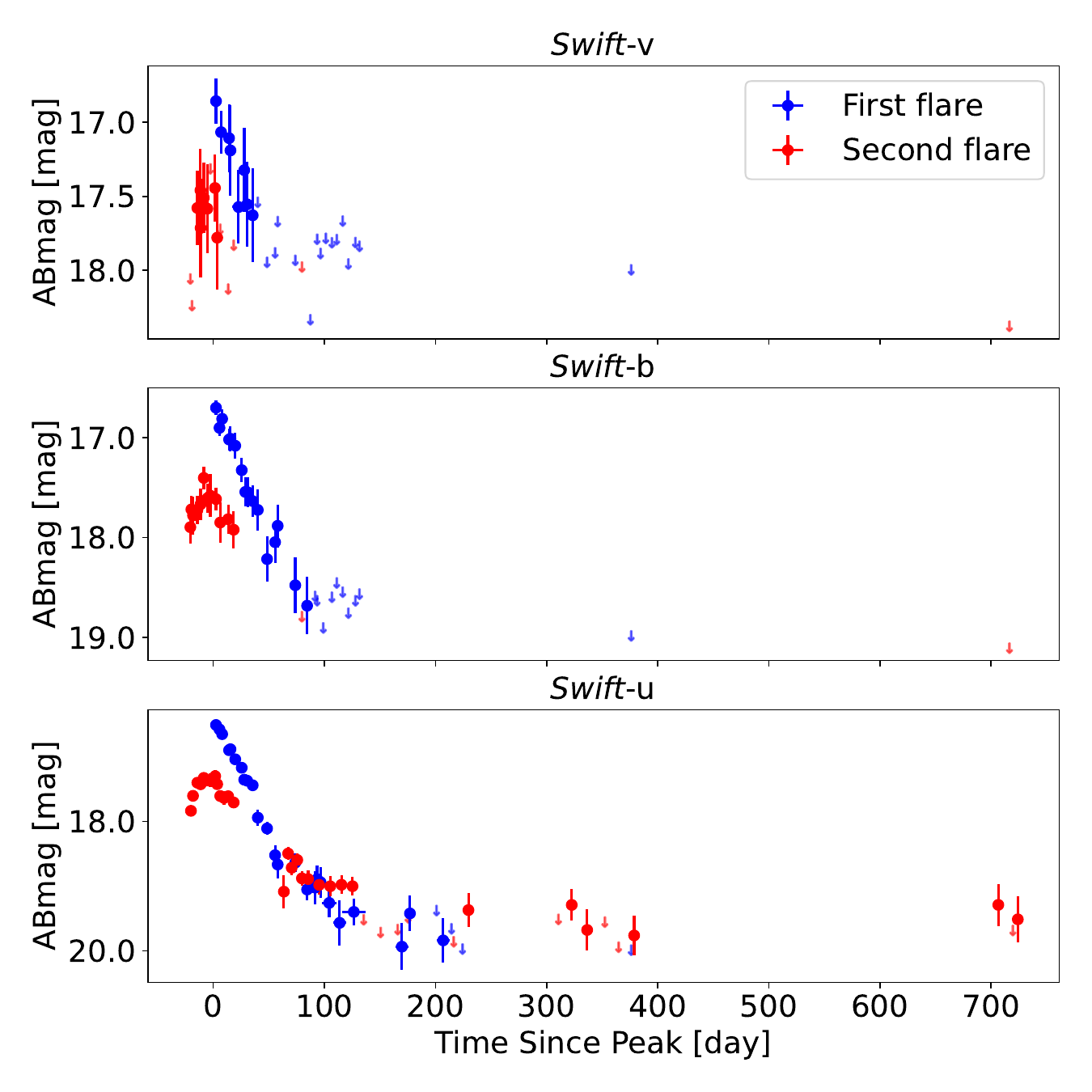}
    \end{minipage}
    \hfill
    \begin{minipage}[b]{0.49\textwidth}
        \centering
        \includegraphics[width=\textwidth]{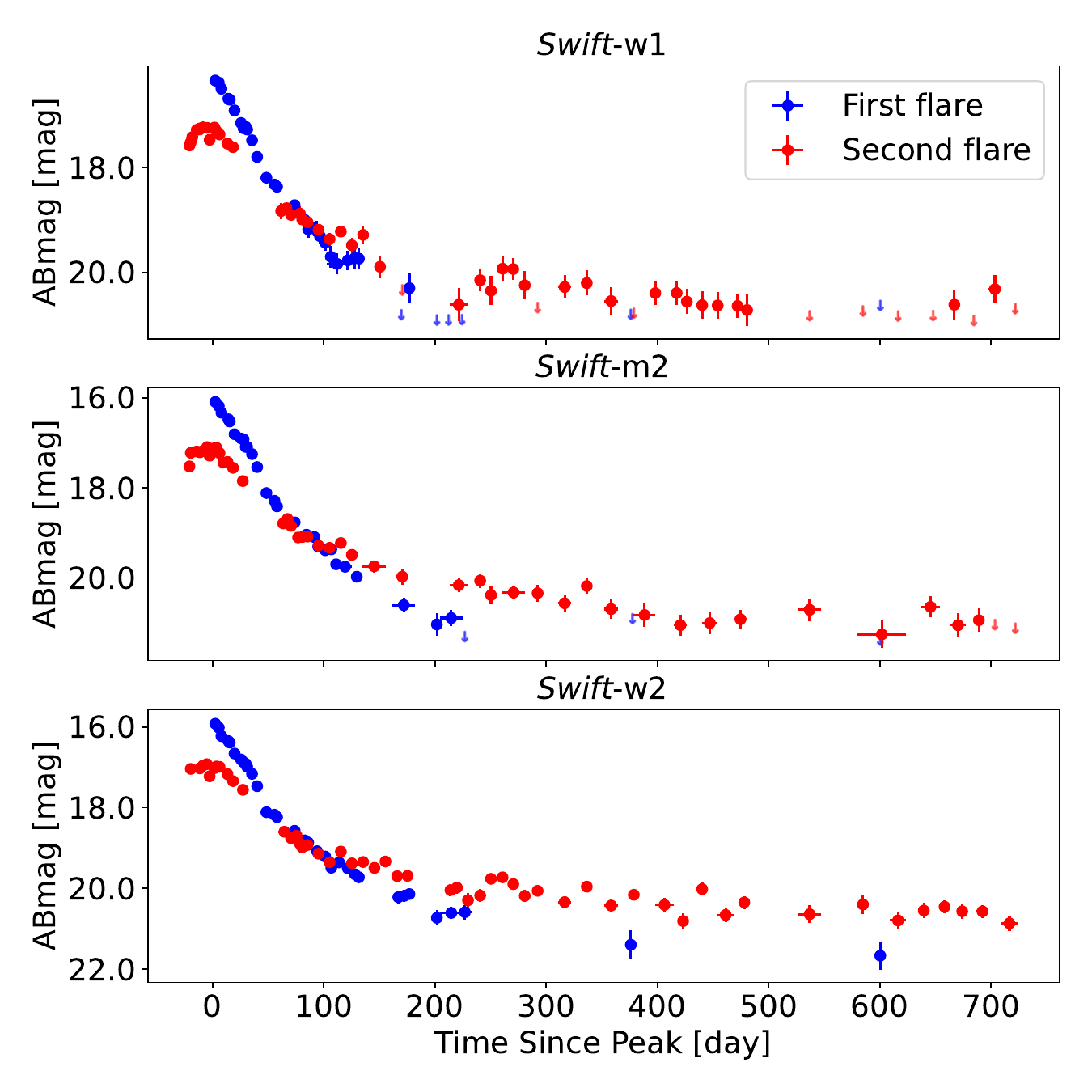}
    \end{minipage}
    \caption{\swift-UVOT light curves of TDE\,2022dbl in the optical bands (left panels) and UV bands (right panels), shown separately for each filter. All photometry is corrected for Galactic extinction and host-galaxy contribution. Downward arrows indicate $3\sigma$ upper limits. The first flare (blue) and second flare (red) are plotted relative to the peak times reported by \citet{makrygianni2025double}.}
    \label{fig:Swift_subplots}
\end{figure*}

Table~\ref{Tab:Extinctions} lists the extinction corrections applied to each photometric band throughout this work. Table~\ref{Tab:Prospector input} provides the photometric inputs used for the \texttt{Prospector} host galaxy modeling described in Section~\ref{sec:hostmodel}, whereas Figure~\ref{fig:corner} shows the corresponding posterior distributions. Tables~\ref{tab:alphacorrection} and~\ref{Tab:PSFcorrection} list the correction factors applied to account for \swift-UVOT instrumental systematics, as described in Section~\ref{SubSec:dataUVOT}. The final \swift-UVOT photometry is provided in Table~\ref{Tab:Swift UVOT}, with the individual filter light curves shown in Figure~\ref{fig:Swift_subplots}. Table~\ref{Tab:speclog} summarizes the general information on optical spectroscopy used in Section~\ref{sec:spectroscopy}. Finally, the best-fit blackbody parameters described in Section~\ref{sec:blackbody} are listed in Table~\ref{Tab: blackbody}. 

\startlongtable
\begin{deluxetable*}{cc}
\tablecaption{Photometry Extinction Corrections  (all values in AB magnitudes)}
\tablehead{
\colhead{Band} & \colhead{Extinction (mag)}
}
\startdata
\swift-$v$ & 0.052 \\
\swift-$b$ & 0.069 \\
\swift-$u$ & 0.083 \\
\swift-uvw1 & 0.112 \\
\swift-uvm2 & 0.154 \\
\swift-uvw2 & 0.151 \\
ZTF-$g$ & 0.062 \\
ZTF-$r$ & 0.040 \\
ZTF-$i$ & 0.030 \\
ATLAS-$o$ & 0.038 \\
ATLAS-$c$ & 0.054 
\enddata
\label{Tab:Extinctions}
\end{deluxetable*}

\startlongtable
\begin{deluxetable*}{ccccc}
\tablecaption{Prospector input Photometry  (all values in AB magnitudes)}
\tablehead{
\colhead{Band} & \colhead{Observed Photometry (mag)} & Extinction (mag) & \colhead{Intrinsic Photometry (mag)} & \colhead{Uncertainty (mag)}
}
\startdata
GALEX\_FUV & 22.424  & 0.133 & 22.291 & 0.2887 \\
GALEX\_NUV & 20.633  & 0.145 & 20.488 & 0.0636 \\
SDSS\_$u$    & 17.852  & 0.080 & 17.772 & 0.0480 \\
SDSS\_$g$    & 16.264  & 0.064 & 16.200 & 0.0055 \\
SDSS\_$r$    & 15.594  & 0.044 & 15.550 & 0.0047 \\
SDSS\_$i$    & 15.256  & 0.032 & 15.224 & 0.0057 \\
SDSS\_$z$    & 15.020  & 0.024 & 14.996 & 0.0161 \\
2MASS\_$J$   & 15.152  & 0.014 & 15.138 & 0.076  \\
2MASS\_$H$   & 14.942  & 0.009 & 14.933 & 0.087  \\
2MASS\_$K$   & 15.021  & 0.006 & 15.015 & 0.106 \\
WISE\_W1   & 16.072  & 0     & 16.072 & 0.026 \\
WISE\_W2   & 16.711  & 0     & 16.711 & 0.030
\enddata
\label{Tab:Prospector input}
\end{deluxetable*}

\startlongtable
\begin{deluxetable*}{ccccc}
\tablecaption{$\alpha$ correction for the \emph{Swift}-UVOT Photometry uncertainty }
\tablehead{
\colhead{Filter} & \colhead{$\alpha$ for mag} & \colhead{Error} & \colhead{$\alpha$ for flux} & \colhead{Error}
}
\startdata
\swift-$v$     & 1.087      & 0.073 & 1.140    &   0.062      \\
\swift-$b$     & 0.915      & 0.092 & 1.10     &   0.17      \\
\swift-$u$     & 1.00        & 0.11  & 1.04    &  0.10       \\
\swift-uvw1     & 1.02      & 0.10  & 1.02    &  0.11       \\
\swift-uvm2     & 0.95      & 0.12  & 0.99    &  0.14       \\
\swift-uvw2     & 1.089      & 0.051 & 1.110   &  0.056    
\enddata
\tablecomments{$\alpha$ is the multiplicative factor applied to the photometric uncertainties reported by the \emph{Swift}/UVOT pipeline.
\label{tab:alphacorrection}}
\end{deluxetable*}

\startlongtable
\begin{deluxetable*}{cccc}
\tablecaption{PSF-corrected Host Galaxy \emph{Swift}-UVOT Photometry (all values in AB magnitudes)}
\tablehead{
\colhead{Filter} & \colhead{Prospector model} & PSF Correction & \colhead{Corrected host}
}
\startdata
$v$    & 15.82      & 0.30            & 16.12                  \\
$b$    & 16.51      & 0.25           & 16.76                  \\
$u$    & 17.77      & 0.46           & 18.23                  \\
uvw1    & 19.54      & 0.19           & 19.73                  \\
uvm2    & 20.40       & 0.18           & 20.57                  \\
uvw2    & 20.84      & 0.00            & 20.84                 
\enddata
\tablecomments{The PSF correction is the offset applied to account for flux lost outside the 10$''$ aperture due to the extended \emph{Swift}/UVOT PSF. Corrected host magnitudes are then used for subtraction.
\label{Tab:PSFcorrection}}
\end{deluxetable*}

\begin{deluxetable*}{ccccc}
\tablecaption{\emph{Swift}-UVOT Photometry of TDE\,2022dbl after correcting for Galactic extinction and subtracting the host galaxy emission. Dates are in MJD. $\Delta t$ represents the time uncertainty for binned measurements and the exposure time for individual observations. All magnitudes is in AB magnitudes.}
\tablehead{
\colhead{MJD} &
\colhead{$\Delta t$} &
\colhead{Filter} &
\colhead{Mag.} &
\colhead{$\sigma_{\rm mag}$} \\
\colhead{(d)} &
\colhead{(d)} &
\colhead{} &
\colhead{(mag)} &
\colhead{(mag)}
}
\startdata
59637.6 & 0.6 & $v$    & 16.86 & 0.15 \\
59637.6 & 0.6 & $b$    & 16.70 & 0.08 \\
59637.6 & 0.6 & $u$    & 16.51 & 0.04 \\
59637.6 & 0.6 & uvw1 & 16.33 & 0.04 \\
59637.6 & 0.6 & uvm2 & 16.09 & 0.04 \\
59637.6 & 0.6 & uvw2 & 15.92 & 0.04 \\
\enddata
\tablecomments{A portion of the table is shown here for guidance regarding its form and content. The complete table is available in machine-readable form.}
\label{Tab:Swift UVOT}
\end{deluxetable*}

\startlongtable
\begin{deluxetable*}{cccccc}
\tablecaption{Optical Spectroscopy}
\tablehead{
\colhead{UT Date} & \colhead{MJD (d)} & Phase (d) & \colhead{Telescope} & \colhead{Instrument} & \colhead{Data Source}
}
\startdata
2022-02-21T14:51:21 & 59631.6 & $-$9.5       & FTN       & FLOYDS     & StarDestroyers \\
2022-03-08T09:18:43 & 59646.4 & 5.2       & LDT       & DeVeny     & UMD \tablenotemark{a}           \\
2022-03-25T08:03:50 & 59663.3 & 22.2       & Shane     & Kast       & TRex \tablenotemark{b}            \\
2022-04-05T08:28:19 & 59674.4 & 33.2       & Shane     & Kast       & TRex           \\
2024-02-05T11:15:21 & 60345.5 & $-$2.9       & Keck      & LRIS       & TRex           \\
2024-02-07T10:27:50 & 60347.4 & $-$0.9       & Keck      & LRIS       & Caltech \tablenotemark{c}       \\
2024-02-14T09:34:33 & 60354.4 & 6.0        & Shane     & Kast       & TRex           \\
2024-03-18T09:11:31 & 60387.4 & 39.0       & Shane     & Kast       & TRex           \\
2024-04-16T10:49:26 & 60416.5 & 68.1       & Shane     & Kast       & TRex           \\
2024-04-30T09:07:11 & 60430.4 & 82.0       & Shane     & Kast       & TRex           \\
2024-06-01T10:32:09 & 60462.4 & 114.1      & Keck      & LRIS       & TRex        
\enddata
\tablenotetext{a}{PI: Cenko.}
\tablenotetext{b}{TRansient EXtragalactic group at UC Berkeley (PIs: Margutti and Chornock).}
\tablenotetext{c}{PI: Ravi.}
\label{Tab:speclog}
\end{deluxetable*}

\startlongtable
\begin{deluxetable*}{ccccc}
\tablecaption{Best-fit Blackbody Parameters}
\tablehead{
\colhead{MJD [day]} & \colhead{Phase [day]} & \colhead{$R_{\rm bb}$ [$10^{14}$ cm]} & \colhead{$T_{\rm bb}$ [$10^4$ K]} & \colhead{$L_{\rm bb}$ [$10^{43}$ erg s$^{-1}$]}}
\startdata
59637.6 & -3.6 & $3.818^{+0.052}_{-0.055}$ & $2.784^{+0.049}_{-0.043}$ & $6.243^{+0.255}_{-0.234}$ \\
59640.7 & -0.5 & $4.061^{+0.047}_{-0.051}$ & $2.585^{+0.036}_{-0.034}$ & $5.242^{+0.161}_{-0.150}$ \\
59643.1 & 1.9 & $4.057^{+0.058}_{-0.051}$ & $2.500^{+0.033}_{-0.035}$ & $4.580^{+0.131}_{-0.131}$ \\
59649.3 & 8.2 & $3.814^{+0.078}_{-0.083}$ & $2.441^{+0.046}_{-0.044}$ & $3.672^{+0.127}_{-0.115}$ \\
59650.6 & 9.4 & $3.804^{+0.088}_{-0.092}$ & $2.408^{+0.049}_{-0.046}$ & $3.467^{+0.122}_{-0.115}$ \\
59654.9 & 13.7 & $3.678^{+0.091}_{-0.090}$ & $2.336^{+0.047}_{-0.048}$ & $2.866^{+0.093}_{-0.099}$ \\
59660.7 & 19.6 & $3.288^{+0.098}_{-0.096}$ & $2.376^{+0.056}_{-0.054}$ & $2.451^{+0.095}_{-0.085}$ \\
59663.0 & 21.8 & $3.154^{+0.089}_{-0.089}$ & $2.385^{+0.055}_{-0.051}$ & $2.296^{+0.085}_{-0.079}$ \\
\enddata
\tablecomments{The first eight rows of the best-fit blackbody parameters are shown here for formatting and guidance. The complete Table can be found in machine-readable form.}
\label{Tab: blackbody}
\end{deluxetable*}

\bibliography{BibliographyFile}{}
\bibliographystyle{aasjournal}



\end{document}